\documentclass[%
 reprint,
 amsmath,amssymb,
 aps,
]{revtex4-1}

\usepackage{graphicx}
\usepackage{dcolumn}
\usepackage{bm}
\usepackage[colorlinks, linkcolor=blue, anchorcolor=blue, citecolor=blue]{hyperref}
\usepackage{url}
\usepackage{amsmath,amssymb}
\usepackage{newtxmath}
\usepackage[mathlines]{lineno}
\usepackage{subfigure}
\usepackage{eurosym}

\begin{document}

\preprint{APS/123-QED}

\title{Coherent single-photon scattering spectra for a giant-atom waveguide-QED system beyond dipole approximation}
\author{Q. Y. Cai}
\affiliation{School of Physical Science and Technology, Southwest Jiaotong University, Chengdu 610031, China}
\author{W. Z. Jia}
\email{wenzjia@swjtu.edu.cn}
\affiliation{School of Physical Science and Technology, Southwest Jiaotong University, Chengdu 610031, China}
\date{\today}

\begin{abstract}
We investigate the single-photon scattering spectra of a giant atom coupled to a one dimensional waveguide via multiple 
connection points or a continuous coupling region. Using a full quantum mechanical method, we obtain the general analytic 
expressions for the single-photon scattering coefficients, which are valid in both the Markovian and the non-Markovian 
regimes. We summarize the influences of the non-dipole effects, mainly caused 
by the phases accumulated by photons traveling between coupling points, on
the scattering spectra. We find that under the Markovian limit, the phase decay is detuning-independent, resulting in Lorentzian 
lineshapes characterized by the Lamb shifts and the effective decay rates. While in the non-Markovian regime, 
the accumulated phases become detuning-dependent, giving rise to non-Lorentzian lineshapes, characterized by multiple side peaks 
and total transmission points. Another interesting phenomenon in the non-Markovian regime is generation of broad photonic band gap 
by a single giant atom. We further generalize the case of discrete coupling points to the continuum 
limit with atom coupling to the waveguide via a continuous area, and  analyze the scattering spectra for some 
typical distributions of coupling strength.

\end{abstract}

\pacs{Valid PACS appear here}
\maketitle


\section{\label{Introduction}Introduction}
It is well known that natural atoms can be looked on as point-like particles for their sizes are much smaller than the wavelength of the light they interact with. 
This treatment, called the dipole approximation, is widely used in quantum optics to simplify the descriptions of the interactions between atoms and photons
\cite{Walls-QuantumOptics}. 
However, recent experiments in the category of waveguide quantum electrodynamics (QED) \cite{Roy-PRM2017, Gu-PhysReports2017} 
have shown that this assumption can be violated \cite{Gustafsson-Science2014}. 
In these setups, superconducting artificial atoms (e.g., transmon qubits \cite{Koch-PRA2007}) can couple to the bosonic modes (surface acoustic waves or microwaves) in a waveguide at multiple points, which are spaced wavelength distances apart, 
form the so called giant-atom structure \cite{Kockum-Arxiv2020, Andersson-Nature2019, Kannan-Nature2020}. 
In waveguide-QED systems containing giant atoms, the multiple coupling points give rise to interference effects, 
resulting in some novel phenomena that are not present in quantum optics with point-like small atoms, 
e.g., frequency-dependent decay rate and Lamb shift of a giant atom \cite{Kannan-Nature2020, Kockum-PRA2014,Vadiraj-PRA2021}, 
and decoherence-free interaction between multiple giant atoms \cite{Kockum-PRL2018,Kannan-Nature2020}. 
By engineering the time delays between coupling points to be comparable to the lifetime of the atom, 
a single giant atom can realize non-Markovian dynamics with polynomial spontaneous decay instead of exponential 
\cite{Guo-PRA2017, Andersson-Nature2019}. 
Another phenomenon for giant atoms in the non-Markovian regime is creation of bound 
states \cite{Guo-PRR2020,Guo-PRA2020}. Besides one dimensional geometries based on superconducting circuits, the giant-atom scheme could be
realized in higher dimensions with cold atoms in optical lattices \cite{Tudela-PRL2019}. 
The interactions between a giant atom and bosonic modes in a nonlinear waveguide or a topological waveguide are also investigated \cite{Zhao-PRA2020, Cheng-Arxiv2021}. 

The strong light-matter interactions in waveguide-QED structures may enable us to control the propagating 
single photons in the waveguide \cite{Roy-PRM2017, Gu-PhysReports2017, Astafiev-Science2010, Zhu-PRA2019, Ask-Arxiv2020}.
On the other hand, the photon scattering spectra can be used to probe 
and characterize the interactions between emitters and photons, including non-dipole effects in giant-atom structures. 
However, to the best of our knowledge, the photon scattering spectra for waveguide-QED 
systems containing a giant atom with multiple coupling points, have not been systematic studied, 
except for some special cases, e.g., a giant atom with two coupling points \cite{Guo-PRA2017}. In this paper, we focus 
on the non-dipole effects on the single-photon transport properties in a waveguide coupled by a giant atom with 
\textit{multiple coupling points} or a \textit{continuous coupling region}. We first obtain the analytical solutions for the
single-photon transmission and reflection coefficients utilizing a real-space scattering method \cite{Shen-PRL2005}. Then we analyze 
the influences of phase delay between coupling points and number of coupling points on the scattering spectra, 
in both the Markovian and the non-Markovian regimes. We mainly focus on the deep non-Markovian regime, where 
the phase-accumulated effects for detuned photons give rise to abundant spectrum structures, including 
non-Lorentzian lineshapes with multiple peaks, broad photonic band gap generated by a single giant atom, and so on. 
The features of spectra characterizing the transition from 
the Markovian to the non-Markovian regime are also discussed. We further generalize the results obtained from the 
case of discrete coupling points to the continuum limit with infinitely many coupling points, or in other words, the 
atom couples to the waveguide through a continuous area. We investigate the influences of 
the effects beyond the dipole approximation on the scattering spectra for some typical distributions of coupling strength. 

It is worth noting that some interesting properties related to photon transport of a giant atom with multiple points \cite{Guo-PRR2020,Guo-PRA2020,Zhao-PRA2020}, including the continuum limit \cite{Guo-PRR2020}, have already been 
investigated before. Here we comment on the connections and differences between these previous works and our work. In Refs.~\cite{Guo-PRR2020} and \cite{Guo-PRA2020}, the relaxation dynamics of an initially excited giant atom with multiple coupling points (including the the continuum limit in Ref.~\cite{Guo-PRR2020}) was studied. The main result is appearance of bound state of bosonic modes in the non-Markovian regime. While in our work, we focus on the single-photon scattering spectra, where the atom is initially in its ground state, and a single photon incidents from the waveguide. 
We also note that in Ref.~\cite{Zhao-PRA2020}, the photon scattering problem for a giant atom coupling to a coupled-resonator array  through multiple connecting points was investigated. In their model, the discrete bosonic modes are provided by a cavity array. On the contrary, our work focuses on waveguide-QED  systems containing a one-dimensional infinite waveguide (e.g. an infinite transmission line) with continuous modes, which are widely adopted in present experiments \cite{Kockum-Arxiv2020, Andersson-Nature2019, Kannan-Nature2020}.

The paper is organized as follows. In Sec.~\ref {Model}, we give a theoretical model, including 
the system Hamiltonian and corresponding equations of motion, and further obtain the transmittance and reflectance of 
single-photons scattering. In Sec.~\ref {ScatteringSpectraDiscrete}, we analyze the scattering spectra for the case 
of discrete coupling points, both in the Markovian and the non-Markovian regimes. In Sec.~\ref {Continuum limit}, 
we discuss the scattering spectra under the continuum limit. Finally, further discussions and conclusions are given 
in Sec.~\ref {conclusion}.
\section{\label{Model}Model and solutions}
\subsection{\label{HamiltonianEQM}Hamiltonian and equations of motion}
\begin{figure}[t]
\includegraphics[width=0.45\textwidth]{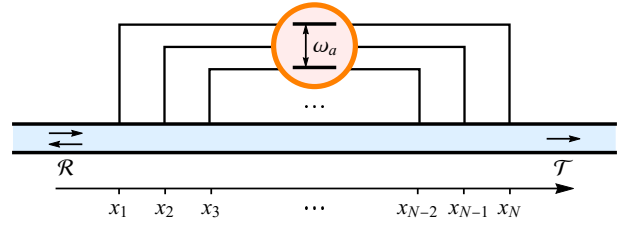}
\caption{A schematic of a giant atom coupled to a one dimensional waveguide at multiple points $x_{m}$.}
\label{system}
\end{figure}

The configuration of the system is shown schematically by Fig.~\ref{system}, a two-level atom directly couples to a one-dimension waveguide at $N$ coupling points. 
Under rotating-wave approximation (RWA), the total Hamiltonian of the system described by Fig.~\ref{system} can be written as ($\hbar=1$)
\begin{eqnarray}
\hat{H}&=&
\omega_\mathrm{a}\left|e\right>\left<e\right|+
\int\mathrm{d}x\hat{c}_\mathrm{R}^{\dagger}\left(x\right)\left(-iv_{\mathrm{g}}\frac{\partial}{\partial x}\right)\hat{c}_\mathrm{R}\left(x\right) 
\nonumber \\
&&+\int\mathrm{d}x\hat{c}_\mathrm{L}^{\dagger}\left(x\right)\left(iv_{\mathrm{g}}\frac{\partial}{\partial x}\right)\hat{c}_\mathrm{L}\left(x\right)
\nonumber \\
&&+\int\mathrm{d}x\sum_{m=1}^{N}V_{m}\delta\left(x-x_{m}\right)\Bigg{[}\sum_{\mathrm{i}=\mathrm{R},\mathrm{L}}\hat{c}_\mathrm{i}^{\dagger}\left(x\right)\sigma_{-}+\mathrm{H}.\mathrm{c}.\Bigg{]}.
\nonumber 
\\
 \label{totalHamiltonian}
 \end{eqnarray}
$\omega_\mathrm{a}$ represents the atomic transition frequency. $\sigma_{+}=|e\rangle\langle g|$ and 
$\sigma_{-}=|g\rangle\langle e|$ are the raising and lowering operators of the atom, 
where $|e\rangle (|g\rangle)$ represents the excited (ground) state. 
$v_\mathrm{g}$ is the group velocity of the photons in the waveguide. 
$\hat{c}_\mathrm{R}(x)$ [$\hat{c}^\dagger_\mathrm{R}(x)$] and $\hat{c}_\mathrm{L}(x)$ [$\hat{c}^\dagger_\mathrm{L}(x)$] 
are the field operators of annihilating (creating) the right- and left-propagating photons at position $x$ in the waveguide. 
$V_{m}$ is the coupling strength between the atom and the waveguide at the coupling point with coordinate $x_{m}$. 

We assume that initially a single photon with energy $E$ incidences.
Thus in the single excitation subspace, the interacting eigenstate of system can be written as
\begin{eqnarray}
\left|\Psi\right>&=&
\int\mathrm{d}x\Phi_\mathrm{R}\left(x\right)\hat{c}^\dagger_\mathrm{R}\left(x\right)\left|\emptyset\right>
\nonumber \\
&&+\int\mathrm{d}x\Phi_\mathrm{L}\left(x\right)\hat{c}^\dagger_\mathrm{L}\left(x\right)\left|\emptyset\right>+f_\mathrm{a}\sigma_{+}\left|\emptyset\right>,
\label{eigenstate}
\end{eqnarray}
where $\Phi_\mathrm{R}(x) [\Phi_\mathrm{L}(x)]$ is the single-photon wave function of a right-moving (left-moving) photon. 
$f_\mathrm{a}$ is the excitation amplitude of the giant atom.  $|\emptyset\rangle$ is the vacuum state, 
which means that there are no photons in the transmission line and the atom is in its ground state. 
Substituting Eq.~\eqref{eigenstate} into the eigen equation
\begin{equation}
\hat{H}\left|\Psi\right>=E\left|\Psi\right>
\label{eigenequation}
\end{equation}
yields the following equations of motion:
\begin{subequations}
	\begin{equation}
	\left(-iv_\mathrm{g}\frac{\partial}{\partial x}-E\right)\Phi_\mathrm{R}\left(x\right)+\sum_{m=1}^{N}V_{m}\delta\left(x-x_{m}\right)f_\mathrm{a}=0,
	\label{EQM1}
	\end{equation}
	\begin{equation}
	\left(iv_\mathrm{g}\frac{\partial}{\partial x}-E\right)\Phi_\mathrm{L}\left(x\right)+\sum_{m=1}^{N}V_{m}\delta\left(x-x_{m}\right)f_\mathrm{a}=0,
	\label{EQM2}
	\end{equation}
	\begin{equation}
	\sum_{m=1}^{N}V_{m}\left[\Phi_\mathrm{R}\left(x_{m}\right)+\Phi_\mathrm{L}\left(x_{m}\right)\right]+\left(\omega_\mathrm{a}-E\right)f_\mathrm{a}=0.
	\label{EQM3}
	\end{equation}
\end{subequations}
\subsection{\label{ScatteringCoefficients}Scattering coefficients}
For a photon incident from the left, $\Phi_\mathrm{R}(x)$ and $\Phi_\mathrm{L}(x)$ take the form
\begin{subequations}
\begin{eqnarray}
\Phi_\mathrm{R}\left(x\right)&=&
e^{ikx}\Big[\vartheta\left(x_{1}-x\right)+\sum_{m=1}^{N-1}t_{m}\vartheta\left(x-x_{m}\right)\vartheta\left(x_{m+1}-x\right)
\nonumber \\
&&+t\vartheta\left(x-x_{N}\right)\Big],
\label{WF1}
\end{eqnarray}
\begin{equation}
\Phi_\mathrm{L}\left(x\right)=
e^{-ikx}\Big[r\vartheta\left(x_{1}-x\right)+\sum_{m=1}^{N-1}r_{m}\vartheta\left(x-x_{m}\right)\vartheta\left(x_{m+1}-x\right)\Big],
\label{WF2}
\end{equation}
\end{subequations}
where $k$ is the wave vector of the photon, $t_m$ ($r_m$) is the transmission (reflection) amplitude for the $m$th [$(m+1)$th] coupling point, 
$t$ ($r$) is the transmission (reflection) amplitude for the last (first) coupling point, 
and $\vartheta(x)$ denotes the Heaviside step function. 

The reflectance and the transmittance can be further defined as $\mathcal{R}=|r|^2$ and $\mathcal{T}=|t|^2$. 
Substituting Eqs.~\eqref{WF1} and \eqref{WF2} into Eqs.~\eqref{EQM1}-\eqref{EQM3}, we can fix $E=v_{\mathrm{g}}k$  and obtain
\begin{widetext}
\begin{subequations}
	\begin{equation}
	\mathcal{T}=\frac{\left(\Delta_{k}-\frac{1}{2}\sum_{m,n=1}^{N}\sqrt{\gamma_{m}\gamma_{n}} \sin{\left|\phi_{mn}\right|}\right)^2}
	{\left(\Delta_{k}-\frac{1}{2}\sum_{m,n=1}^{N}\sqrt{\gamma_{m}\gamma_{n}}\sin{\left|\phi_{mn}\right|}\right)^2
	+\frac{1}{4}\left(\sum_{m,n=1}^{N}\sqrt{\gamma_{m}\gamma_{n}}\cos{\phi_{mn}}\right)^2},
	\label{transmission}
	\end{equation}
	\begin{equation}
	\mathcal{R}=\frac{\frac{1}{4}\left(\sum_{m,n=1}^{N}\sqrt{\gamma_{m}\gamma_{n}}\cos{\phi_{mn}}\right)^2}
	{\left(\Delta_{k}-\frac{1}{2}\sum_{m,n=1}^{N}\sqrt{\gamma_{m}\gamma_{n}}\sin{\left|\phi_{mn}\right|}\right)^2
	+\frac{1}{4}\left(\sum_{m,n=1}^{N}\sqrt{\gamma_{m}\gamma_{n}}\cos{\phi_{mn}}\right)^2},
	\label{reflection}
\end{equation}
\end{subequations}
\end{widetext}
where $\Delta_{k}=v_\mathrm{g}k-\omega_\mathrm{a}$ is the detuning between the single photon and the atom. 
Besides, we define the decay rate of single coupling point $\gamma_{m}=2V_{m}^{2}/v_\mathrm{g}$ and 
the phase delay $\phi_{mn}=k(x_{m}-x_{n})=(1+\Delta_{k}/\omega_\mathrm{a})\tilde{\phi}_{mn}$, where $\tilde{\phi}_{mn}=\omega_\mathrm{a}(x_{m}-x_{n})/v_\mathrm{g}$. 
Note that the conservation of photon number results in  $\mathcal{T}+\mathcal{R}=1$. 

It should be emphasized 
that the expressions \eqref{transmission} and \eqref{reflection} are valid in both the Markovian and the non-Markovian regimes.
In a single giant atom, non-Markovianity can be realized by engineering the time delays between coupling points to be comparable to the relaxation time \cite{Guo-PRA2017, Ask-PRA2019}.
Other systems that can exhibit time-delay effects include an atom in front of a mirror \cite{Dorner-PRA2002, Tufarelli-PRA2013, Pichler-PRL2016, Guimond-PRA2016} and distant atoms coupled locally to the same environment \cite{Laakso-PRL2014, Fang-PRA2015, Pichler-PRL2016, Milonni-PRA1974, Rist-PRA2008, Loo-Science2013, Fang-RPJ2014}.
For a giant atom with multiple coupling points spaced, we can define a characteristic time $T=(x_N-x_1)/v_\mathrm{g}=\tilde{\phi}_{N1}/\omega_{\mathrm{a}}$ for light to travel between the leftmost and the rightmost coupling points. 
In the Markovian regime, $T$ should be much smaller than the relaxation time $1/\tilde{\Gamma}$, where 
\begin{equation}
\tilde{\Gamma}=\left(\sum_{m=1}^{N}\sqrt{\gamma_{m}}\right)^2
\label{DecayDipole}
\end{equation}
is the atomic decay rate in the dipole-approximation limit $\tilde{\phi}_{N1}\to0$. When the condition $T\sim1/\tilde{\Gamma}$ is satisfied,  the giant atom enters the non-Markovian regime. 
Alternatively, we can expressed the non-Markovian condition in terms of phase delay as
\begin{equation}
\tilde{\phi}_{N1}\sim\frac{\omega_{\mathrm{a}}}{\tilde{\Gamma}}.
\label{nonMarkovianCondition1}
\end{equation}

Under the Markovian limit, we can replace the detuning-dependent phase factor $\phi_{mn}$ in Eqs.~\eqref {transmission} and \eqref {reflection}  by the detuning-independent one
$\tilde{\phi}_{mn}$. Such an approximation requires $\tilde{\phi}_{mn}\Delta_{k}/\omega_\mathrm{a}\ll1$, and this is true under Markovian limit 
$\tilde{\phi}_{N1}\ll\omega_{\mathrm{a}}/\tilde{\Gamma}$, given that the bandwidth $\Delta_{k}$ of interest here is of the order of $\tilde{\Gamma}$. 
Then the corresponding transmission and reflection coefficients become
\begin{subequations}
	\begin{equation}
	\mathcal{T}=\frac{\left(\Delta_{k}-\Delta_\mathrm{L}\right)^2}{\left(\Delta_{k}-\Delta_\mathrm{L}\right)^2+\frac{1}{4}\Gamma_\mathrm{eff}^2},
	\label{transmissionMarkov}
	\end{equation}
	
	\begin{equation}
	\mathcal{R}=\frac{\frac{1}{4}\Gamma_\mathrm{eff}^2}{\left(\Delta_{k}-\Delta_\mathrm{L}\right)^2+\frac{1}{4}\Gamma_\mathrm{eff}^2},
	\label{reflectionMarkov}
	\end{equation}
\end{subequations}
giving rise to Lorentzian lineshapes centered at $\Delta_{k}=\Delta_{\mathrm{L}}$ with width $\Gamma_{\mathrm{eff}}$, where
\begin{subequations}
\begin{equation}
\Delta_{\mathrm{L}}=\frac{1}{2}\sum_{m,n=1}^{N}\sqrt{\gamma_{m}\gamma_{n}} \sin{\left|\tilde{\phi}_{mn}\right|},
\label{lambshift}
\end{equation}
\begin{equation}
\Gamma_{\mathrm{eff}}=\sum_{m,n=1}^{N}\sqrt{\gamma_{m}\gamma_{n}}\cos{\tilde{\phi}_{mn}}
\label{gammaeff}
\end{equation}
\end{subequations}
are the Lamb shift and effective decay rate, respectively \cite{Kockum-PRA2014}. The Lamb shift is caused by the 
interaction with the vacuum fluctuations of the bosonic field and the effective decay rate is 
relevant to the interference effects of re-emitted photons from different coupling points. 

Based on these results, 
we will investigate the single-photon transport properties for the case of discrete coupling points in next section. 
And we will further discuss the continuum limit with infinitely many coupling points in Sec.~\ref{Continuum limit}. 
\section{\label{ScatteringSpectraDiscrete}Scattering spectra for the case of discrete coupling points}
Here we focus on the case that the distance between neighboring connection points is a constant $d$ and the 
coupling strength is the same at each coupling point (with $\gamma_{m}=\gamma$). By defining the detuning-independent 
and detuning-dependent phase delays between neighboring 
points as $\tilde{\theta}=\omega_\mathrm{a}d/v_\mathrm{g}$ and $\theta=kd=(1+\Delta_{k}/\omega_\mathrm{a})\tilde{\theta}$,
one can write ${\phi}_{mn}$ as ${\phi}_{mn}=(m-n){\theta}$.
Correspondingly, the transmission and reflection coefficients given by Eqs.~\eqref{transmission} and \eqref{reflection} become
\begin{subequations}
	\begin{equation}
	\mathcal{T}=\frac{\left(\Delta_{k}-\frac{\gamma}{2}\frac{N\sin\theta-\sin N\theta}{1-\cos\theta}\right)^2}
	{\left(\Delta_{k}-\frac{\gamma}{2}\frac{N\sin\theta-\sin N\theta}{1-\cos\theta}\right)^2+\frac{\gamma^2}{4}\left(\frac{1-\cos N\theta}{1-\cos\theta}\right)^2},
	\label{transmissionMS}
	\end{equation}	
	\begin{equation}
	\mathcal{R}=\frac{\frac{\gamma^2}{4}\left(\frac{1-\cos N\theta}{1-\cos\theta}\right)^2}
	{\left(\Delta_{k}-\frac{\gamma}{2}\frac{N\sin\theta-\sin N\theta}{1-\cos\theta}\right)^2+\frac{\gamma^2}{4}\left(\frac{1-\cos N\theta}{1-\cos\theta}\right)^2}.
	\label{reflectionMS}
	\end{equation}
\end{subequations}
Note that presently the phase delay between $x_N$ and $x_1$ is $\tilde{\phi}_{N1}=(N-1)\tilde{\theta}$, and the atomic decay rate 
under the dipole-approximation limit becomes  $\tilde{\Gamma}=N^2\gamma$. Thus the non-Markovian condition 
\eqref{nonMarkovianCondition1} can be rewritten as
\begin{equation}
N^2(N-1)\tilde{\theta}\sim\frac{\omega_\mathrm{a}}{\gamma}.
\label{nonMarkovianCondition2}
\end{equation}
In the following we will investigate the features of scattering spectra in the Markovian and the non-Markovian regimes, respectively. 
\subsection{\label{SpectraDiscreteMarkov}Spectra for the discrete case: the Markovian regime }
\begin{figure}[t]
	\includegraphics[width=0.45\textwidth]{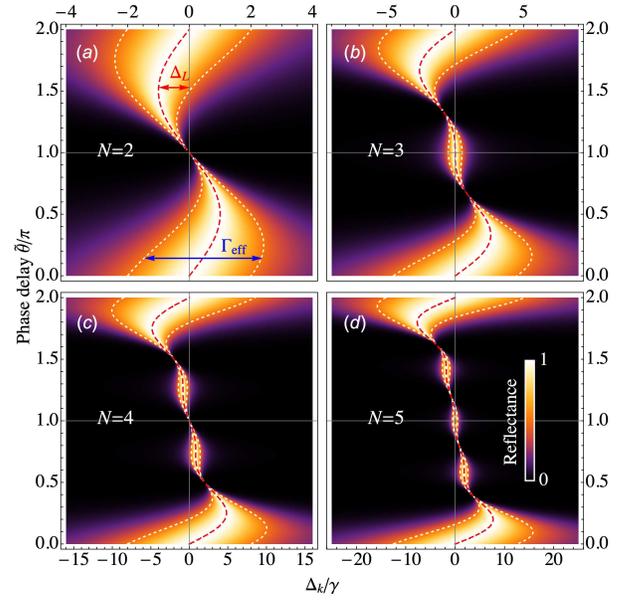}
	\caption{Reflectances $\mathcal{R}$ as functions of the scale detuning $\Delta_{k}/\gamma$ and the phase delay $\tilde{\theta}$ for different $N$:
	(a) $N=2$, (b) $N=3$, (c) $N=4$, (d) $N=5$. The red long-dashed lines are used to label the locations of the reflection peaks, i.e., the Lamb shifts $\Delta_{\mathrm{L}}$. 
	The distances between the white short-dashed lines equal to the effective decay rates $\Gamma_{\mathrm{eff}}$.}
	\label{discretemarkov}
\end{figure}

Under the Markov limit, we can replace the phase factor $\theta$ in Eqs.~\eqref {transmissionMS} and \eqref {reflectionMS}  by 
$\tilde{\theta}$. Then the corresponding transmission and reflection coefficients can be expressed by Eqs.~\eqref{transmissionMarkov} and ~\eqref{reflectionMarkov},
with Lamb shift and effective decay rate 
taking the form
\begin{subequations}
\begin{equation}
\Delta_{\mathrm{L}}=\frac{\gamma}{2}\frac{N\sin\tilde{\theta}-\sin N\tilde{\theta}}{1-\cos\tilde{\theta}},
\label{lambshiftD}
\end{equation}
\begin{equation}
\Gamma_{\mathrm{eff}}=\gamma\frac{1-\cos N\tilde{\theta}}{1-\cos\tilde{\theta}}.
\label{gammaeffD}
\end{equation}
\end{subequations}

Now we begin to discuss the scattering spectra for this case. Without loss of generality, in the following we focus on the reflectance 
$\mathcal{R}$ only, for the transmittance $\mathcal{T}$ and the reflectance $\mathcal{R}$ are constrained by the relation $\mathcal{T}+\mathcal{R}=1$. 
In Fig.~\ref{discretemarkov}, we plot the reflectance $\mathcal{R}$ as functions of the scaled detuning 
$\Delta_{k}/\gamma$ and the phase delay $\tilde{\theta}$ for different $N$. Note that $\Delta_{\mathrm{L}}$ and $\Gamma_{\mathrm{eff}}$ 
changes periodically with $\tilde{\theta}$, thus without loss of generality, in Fig.~\ref{discretemarkov} the range of $\tilde{\theta}$ is chosen as $\tilde{\theta}\in[0,2\pi]$. 
In the Markovian regime considered here, for different phase delay $\tilde{\theta}$ and number of coupling points $N$, 
the reflection spectra have a Lorentzian line shape centered at $\Delta_{k}=\Delta_{\mathrm{L}}$ with width $\Gamma_{\mathrm{eff}}$ 
(labeled by the red long-dashed lines and white short-dashed lines in Fig.~\ref{discretemarkov}, respectively). 
We can see that the Lamb shift $\Delta_{\mathrm{L}}=0$ when $\tilde\theta=n\pi$ ($n\in\mathbb{N}$). 
The effective decay $\Gamma_{\mathrm{eff}}$ reach maximum value $\Gamma_{\mathrm{eff}}=\tilde{\Gamma}$ at $ \tilde{\theta}=2n\pi$  ($n\in\mathbb{N}$). 
Besides, when $N>2$ [see Figs.~\ref{discretemarkov}(b)-\ref{discretemarkov}(d)], the effective decay $\Gamma_{\mathrm{eff}}$ takes some additional local maximum values. 
From Eq.~\eqref{gammaeffD}, one can derive that the condition for reaching these values is $N\cot(N\tilde{\theta}/2)=\cot(\tilde{\theta}/2)$. Finally, 
when $\tilde{\theta}=2n\pi/N$ ($n\in\mathbb{N}^{+}$, and $\mathrm{mod}\left[n,N\right]\neq0$), the effective decay $\Gamma_{\mathrm{eff}}=0$, 
i.e., the atom decouples from the waveguide, resulting in total transmission of incident photons [with $\mathcal{R}=0$ for all $\Delta_{k}$, see Fig.~\ref{discretemarkov}]. 
\subsection{\label{SpectraDiscreteNonMarkov}Spectra for the discrete case: the non-Markovian regime}
\begin{figure}[t]
	\includegraphics[width=0.5\textwidth]{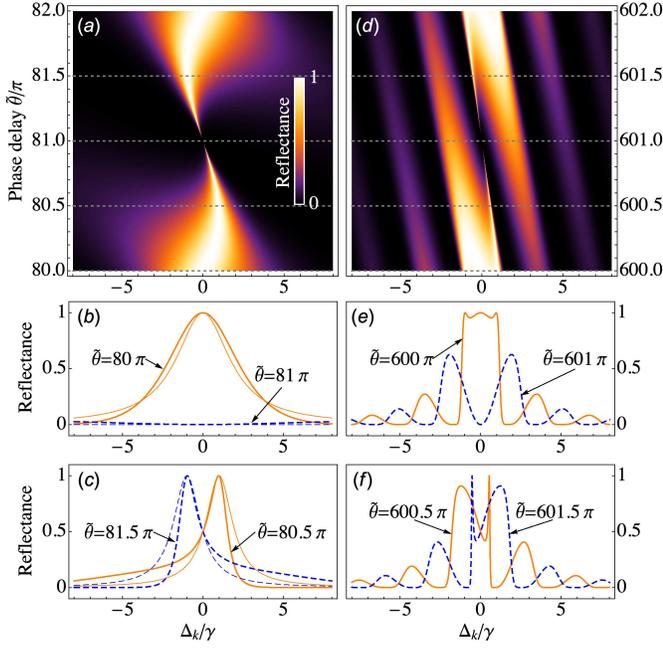}
	\caption{Reflectances in the moderately non-Markovian regime [(a)-(c)] and the deep non-Markovian regime [(d)-(f)] for fixed $N=2$ but different 
	$\tilde{\theta}$. Here we set $\omega_\mathrm{a}=10^3\gamma$. 
	(a) Reflectances $\mathcal{R}$ as functions of the scaled detuning $\Delta_{k}/\gamma$ and the phase delay $\tilde{\theta}$ 
	in the moderately non-Markovian 
	regime. The orange solid line and blue dashed line show cross sections at the phase delays $\tilde{\theta}=80\pi$ and $\tilde{\theta}=81\pi$ for (b) and  
	$\tilde{\theta}=80.5\pi$ and $\tilde{\theta}=81.5\pi$ for (c) respectively, which are indicated by the dashed lines in the panel (a). 
	The solid and dashed thin lines in (b) and (c) are plotted according to the expressions 
	Eq~\eqref{reflectionMarkov} under the Markovian limit for comparison.  
	(d) Reflectances $\mathcal{R}$ as functions of the scaled detuning $\Delta_{k}/\gamma$ and the phase delay $\tilde{\theta}$ in the deep 
	non-Markovian regime. 
	The orange solid line and blue dashed line show cross sections at the phase delays $\tilde{\theta}=600\pi$ and $\tilde{\theta}=601\pi$ for (e) and  
	$\tilde{\theta}=600.5\pi$ and $\tilde{\theta}=601.5\pi$ for (f) respectively, which are indicated by the dashed lines in the panel (d). }
	\label{discretenonmarkov1}
\end{figure}
\begin{figure}[t]
	\includegraphics[width=0.5\textwidth]{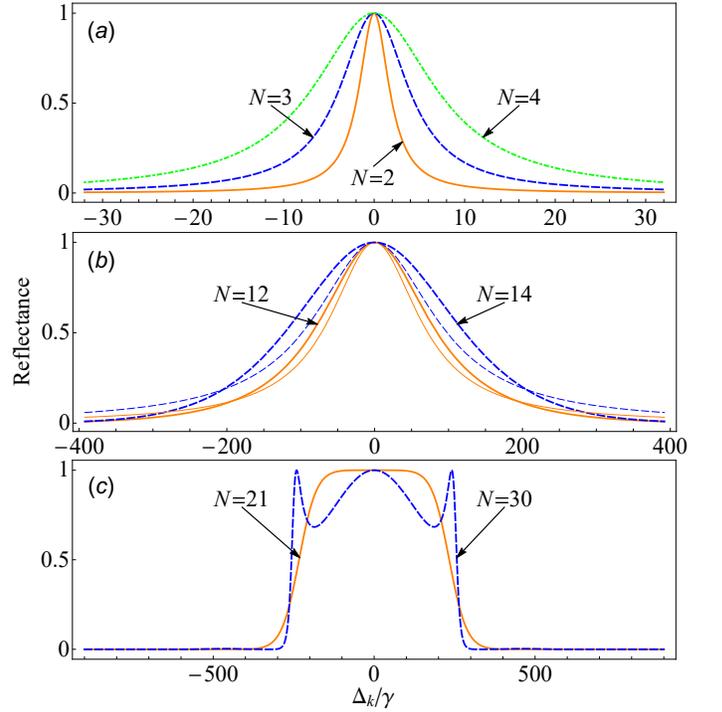}
	\caption{Reflection spectra for fixed $\tilde{\theta}=2\pi$ and different $N$. Here we set $\omega_\mathrm{a}=10^4\gamma$. 
		(a) Reflectances in the Markovian regime with $N=2$ (orange solid line), $N=3$ (blue dashed line) and $N=4$ (green dot-dashed line), respectively. 
		(b) Reflectances in the moderately non-Markovian regime with $N=12$ (orange solid line) and $N=14$ (blue dashed line).  
		The curves with Lorentzian lineshapes (solid and dashed thin lines) are plotted according to the expressions 
	        Eq~\eqref{reflectionMarkov} under the Markovian limit for comparison. 
		(c) Reflectances in the deep non-Markovian regime with $N=21$ (orange solid line) and $N=30$ (blue dashed line).}
	\label{discretenonmarkov2}
\end{figure}
In the non-Markovian regime, the phase $\tilde{\theta}\Delta_{k}/\omega_\mathrm{a}$ accumulated in free propagation for finite detuning cannot be neglected. 
Thus Eqs.~\eqref{transmissionMS} and \eqref{reflectionMS} with detuning-dependent phase $\theta=(1+\Delta_{k}/\omega_\mathrm{a})\tilde{\theta}$ 
should be used to describe the transmission and reflection property of the system. 
In addition, one can see from Eq.~\eqref{nonMarkovianCondition2} that the system can enter the non-Markovian regime 
by increasing either the phase delay $\tilde{\theta}$ or the number of coupling points $N$. 
In the following we will discuss these two cases, respectively. 
\subsubsection{Influences of $\tilde{\theta}$: non-Lorentzian lineshapes due to detuning-dependent phase-accumulated effects}
The reflection spectra in the non-Markovian regime for fixed $N=2$ but different $\tilde{\theta}$ are shown in Fig.~\ref{discretenonmarkov1}. 
Specifically, the spectra in the moderately non-Markovian regime (with $T\sim1/\tilde{\Gamma}$, 
or equally $\tilde{\phi}_{N1}\sim\omega_{\mathrm{a}}/\tilde{\Gamma}$) 
are shown in Figs.~\ref{discretenonmarkov1}(a)-\ref{discretenonmarkov1}(c) and the spectra in the
deep non-Markovian regime (with $T\gg1/\tilde{\Gamma}$, or equally $\tilde{\phi}_{N1}\gg\omega_{\mathrm{a}}/\tilde{\Gamma}$) 
are shown in Figs.~\ref{discretenonmarkov1}(d)-\ref{discretenonmarkov1}(f). 
We can see that in both the moderately and the deep non-Markovian regimes, the lineshapes are non-Lorentzian, 
and in addition the spectra become asymmetric when $\tilde{\theta}\neq n\pi$, which is very different from those in the 
Markovian regime. 

In the moderately non-Markovian regime, the curves begin to deviate from the Lorentzian lineshape, but still remain the single-peak structure,     
as shown in Figs.~\ref{discretenonmarkov1}(b) and~\ref{discretenonmarkov1}(c). For comparison, we also plot the spectra under the Markovian limit, 
as shown by the solid and dashed thin lines in Figs.~\ref{discretenonmarkov1}(b) and~\ref{discretenonmarkov1}(c).

In the deep non-Markovian regime, the spectra exhibit more abundant structures, as shown in Figs.~\ref {discretenonmarkov1}(d)-\ref {discretenonmarkov1}(f). 
Firstly, when $\tilde{\theta}=2n\pi$, and 
\begin{equation}
\tilde{\theta}>\frac{6\omega_\mathrm{a}}{N(N^2-1)\gamma},
\label{3fullreflection}
\end{equation}
there appears two addition total reflection points in the central peak, as shown by the orange solid line in Fig.~\ref{discretenonmarkov1}(e). 
Otherwise, there is only one total reflection point at $\Delta_{k}=0$. In addition, when the phase delay is taken as $\tilde{\theta}=(2n+1)\pi$, 
different from the vanished reflection spectra in the Markovian regime 
[or almost vanished in the moderately non-Markovian regime, see dashed thick lines (blue) in Fig.~\ref{discretenonmarkov1}(b)], 
there appears some reflection peaks at $\Delta_k\neq 0$, because the phase accumulated in free propagation for finite detuning cannot be neglected. 
Finally, the phase-accumulated effects for detuned photons can give rise to some total-transmission points in the spectra, 
as shown in Figs.~\ref{discretenonmarkov1}(d)-\ref{discretenonmarkov1}(f). Specifically, according to Eq.~\eqref {reflectionMS}, 
the condition for total transmission, $\mathcal{R}=0$, is $\cos N\theta-1=0$ and $\cos\theta-1\neq 1$. 
By using this condition and letting $\tilde{\theta}=2n\pi+\tilde{\delta}$ 
($\tilde{\delta}\in\left[0,2\pi\right), n\in\mathbb{N}$), the locations of the total-transmission points shown 
in Figs.~\ref{discretenonmarkov1}(d)-\ref{discretenonmarkov1}(f) can be expressed as  
\begin{equation}
\Delta_{k}=\left(\frac{2n'\pi}{N}-\tilde{\delta}\right)\frac{\omega_\mathrm{a}}{\tilde{\theta}},
\label{fulltransmission}
\end{equation}
where $n'\in\mathbb{Z}$ and$\mod\left[n',N\right]\neq0$. This means that the atom decouples from the photon modes with these frequencies, 
due to destructive interference. 
\subsubsection{Influences of $N$: generation of photonic band gap}
Next we turn to investigate the influence of the numbers of coupling points $N$. The reflection spectra for fixed $\tilde{\theta}=2\pi$ and different $N$ 
are shown in Fig.~\ref{discretenonmarkov2}. For small $N$ satisfying the Markov condition $N^2\left(N-1\right)\tilde{\theta}\ll\omega_\mathrm{a}/\gamma$, 
the reflectance has a Lorentzian lineshape with the width $\tilde{\Gamma}=N^2\gamma$, as shown in Fig.~\ref{discretenonmarkov2}(a). 
For $N$ satisfying the moderately non-Markov condition $N^2\left(N-1\right)\tilde{\theta}\sim\omega_\mathrm{a}/\gamma$, 
the reflection spectra are shown by the orange solid and blue dashed line in Fig.~\ref{discretenonmarkov2}(b). For comparison, we also plot the Lorentzian spectra under the Markovian limit 
using Eq.~\ref{reflectionMarkov}, as shown by the solid and dashed thin lines in Fig.~\ref{discretenonmarkov2}(b). 
One can find that the reflection spectra in the moderately non-Markovian regime slightly deviate from the Lorentzian line shape. 
Besides, with the number of coupling points increasing, the width of peak, which can also be approximated as $\tilde{\Gamma}=N^2\gamma$, will increase. 

Now we consider the reflection spectra for the case of $N$ satisfying the deep Markov condition 
$N^2\left(N-1\right)\tilde{\theta}\gg\omega_\mathrm{a}/\gamma$ (Note that $N$ should be carefully chosen to satisfy $\tilde{\Gamma}\ll\omega_\mathrm{a}$ 
to ensure the validity of the RWA.).  If the value of $N$ reaches the threshold for appearing three maximums in the central peak 
[given by Eq.~\eqref{3fullreflection}], the spectrum exhibits a band gap forbidding transmission of photons, as shown by the orange solid line 
in Fig.~\ref {discretenonmarkov2}(c). 
The width of the band gap has an order of magnitude of $\tilde{\Gamma}=N^2\gamma$. Thus for solid-state quantum system 
(usually with $\gamma/\omega_{\mathrm{a}}\simeq10^{-4}\sim10^{-3}$ \cite{Astafiev-Science2010, Hoi-PRL2011}), 
only a small amount of coupling points are required to obtain a broad band width up to 
$0.1\omega_{\mathrm{a}}$. Note that a photonic band gap can also generated by a waveguide-QED system with an $N$-qubit 
array \cite{Chang-NJP2012, Greenberg-PRA2021}. 
However, for waveguide-QED system with giant atoms, only a \textit{single} atom with multiple coupling points is required to achieve this goal, 
which may greatly decrease the complexity of the design. In combination with tunable couplings between artificial atoms 
(e.g., superconducting qubits) and waveguides \cite{Gu-PhysReports2017}, 
this property could be used as broadband single-photon switches. Finally, when $N$ is further increased, 
becoming larger than the threshold given by Eq.~\eqref{3fullreflection}, there appears three total reflection points in the central peak 
instead one [see the blue dashed line in Fig.~\ref{discretenonmarkov2}(c)]. 
Another feature of the spectra in this regime is the steep walls on both sides of the central peak with a distance about $2\omega_\mathrm{a}/N$.   
\section{\label{Continuum limit}Scattering spectra under the continuum limit}
\begin{figure}[t]
\includegraphics[width=0.45\textwidth]{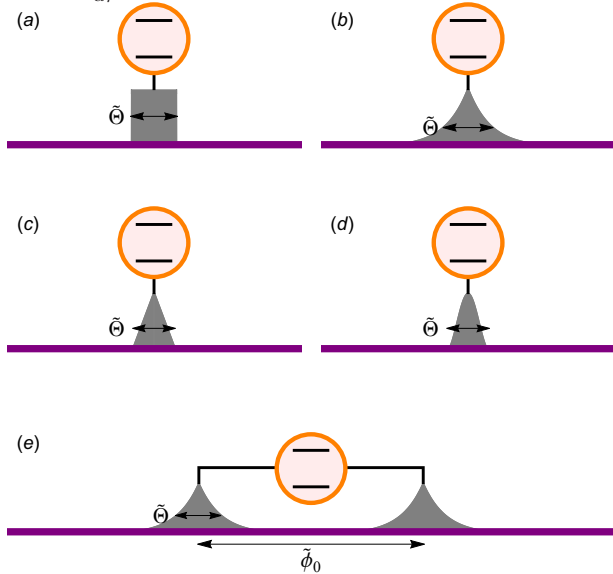}
\caption{Schematics of an atom coupling to a waveguide via a region with some typical coupling distributions: (a) uniform distribution, 
         (b) exponential distribution, (c) triangular distribution, 
	(d) raised cosine distribution, (e) double-exponential distribution.}
\label{continuumlimit}
\end{figure}
In previous discussions, each coupling area is considered as point-like. 
Here we will generalize the results obtained from discrete-coupling configurations to the continuum limit. 
Namely, the coupling area can be regarded as an ensemble of infinitely many coupling points with coupling strengths described by a distribution function. 
These configurations can be more easily realized by giant-atom waveguide-QED system based on superconducting circuits, 
where the distance between coupling points can be designed to be much smaller than the wave length of microwave photons. 
Under the continuum limit, Eqs.~\eqref{transmission}-\eqref{reflection} become
\begin{widetext}
\begin{subequations}
	\begin{equation}
	\mathcal{T}=\frac{\left[\Delta_{k}-\int\mathrm{d}\tilde{\phi}\mathrm{d}\tilde{\phi}'v\left(\tilde{\phi}\right)v\left(\tilde{\phi}'\right)\sin\left|\phi-\phi'\right|\right]^2}
	{\left[\Delta_{k}-\int\mathrm{d}\tilde{\phi}\mathrm{d}\tilde{\phi}'v\left(\tilde{\phi}\right)v\left(\tilde{\phi}'\right)\sin\left|\phi-\phi'\right|\right]^2+
	\left[\int\mathrm{d}\tilde{\phi}\mathrm{d}\tilde{\phi}'v\left(\tilde{\phi}\right)v\left(\tilde{\phi}'\right)\cos\left(\phi-\phi'\right)\right]^2},
	\label{transmissionc}
	\end{equation}
	\begin{equation}
	\mathcal{R}=\frac{\left[\int\mathrm{d}\tilde{\phi}\mathrm{d}\tilde{\phi}'v\left(\tilde{\phi}\right)v\left(\tilde{\phi}'\right)\cos\left(\phi-\phi'\right)\right]^2}
	{\left[\Delta_{k}-\int\mathrm{d}\tilde{\phi}\mathrm{d}\tilde{\phi}'v\left(\tilde{\phi}\right)v\left(\tilde{\phi}'\right)\sin\left|\phi-\phi'\right|\right]^2+
		\left[\int\mathrm{d}\tilde{\phi}\mathrm{d}\tilde{\phi}'v\left(\tilde{\phi}\right)v\left(\tilde{\phi}'\right)\cos\left(\phi-\phi'\right)\right]^2},
	\label{reflectionc}
	\end{equation}
\end{subequations}
\end{widetext}
where $v(\tilde{\phi})$ is the distribution of coupling strength, satisfying $\int\mathrm{d}\tilde{\phi}v\left(\tilde{\phi}\right)=(\tilde{\Gamma}/2)^{1/2}$. 
The phase is defined as $\phi=kx=(1+\Delta_{k}/\omega_\mathrm{a})\tilde{\phi}$, where $\tilde{\phi}=\omega_\mathrm{a}x/v_\mathrm{g}$. 
And $\tilde{\Gamma}$ is, like the discrete case, the atomic decay rate under the dipole-approximation limit, 
where all the couplings are regarded as converging to a single point. Note that here we achieve the scattering
amplitudes under the continuum limit by directly using the discrete expressions Eqs.~\eqref{transmission}-\eqref{reflection}.  One can also  
at first provide the continuous forms of the Hamiltonian and the wave functions, and then solve the corresponding eigen equations to obtain the 
scattering amplitudes (see Appendix~\ref{ContinuumHamiltonian}). 

In the following discussion, we focus on these typical distributions: (a) uniform distribution, (b) exponential distribution, 
(c) triangular distribution, (d) raised cosine distribution, (e) double-exponential distribution, as shown schematically in Fig.~\ref{continuumlimit}. 
For comparison purposes, we let all distributions give rise to the same $\tilde{\Gamma}$.
The expressions of these distribution functions are given in Appendix~\ref{CouplingDistribution}.  
For cases (a)-(d), the coupling strength is concentrated around one point with characteristic width $\tilde{\Theta}$, 
and for case (e) there are two coupling-concentrated areas with a distance $\tilde{\phi}_0$ (Here we express the
width/distance in terms of phase delay for photons with wave vector $\omega_{\mathrm{a}}/v_{\mathrm{g}}$.). Accordingly, 
the characteristic time scale for light traveling through the coupling area is $T=\tilde{\Theta}/\omega_{\mathrm{a}}$ 
[$T=(\tilde{\Theta}+\tilde{\phi_0})/\omega_{\mathrm{a}}$] for cases (a)-(d) [for case (e)].  
Bellow we will discuss the features of scattering spectra for the continuum limit in both the Markovian 
(with $T\ll1/\tilde{\Gamma}$) and the non-Markovian regimes (with $T>1/\tilde{\Gamma}$).
 \subsection{\label{SpectraContinuumMarkov}Spectra under the continuum limit: the Markovian regime}
\begin{figure}[t]
	\includegraphics[width=0.5\textwidth]{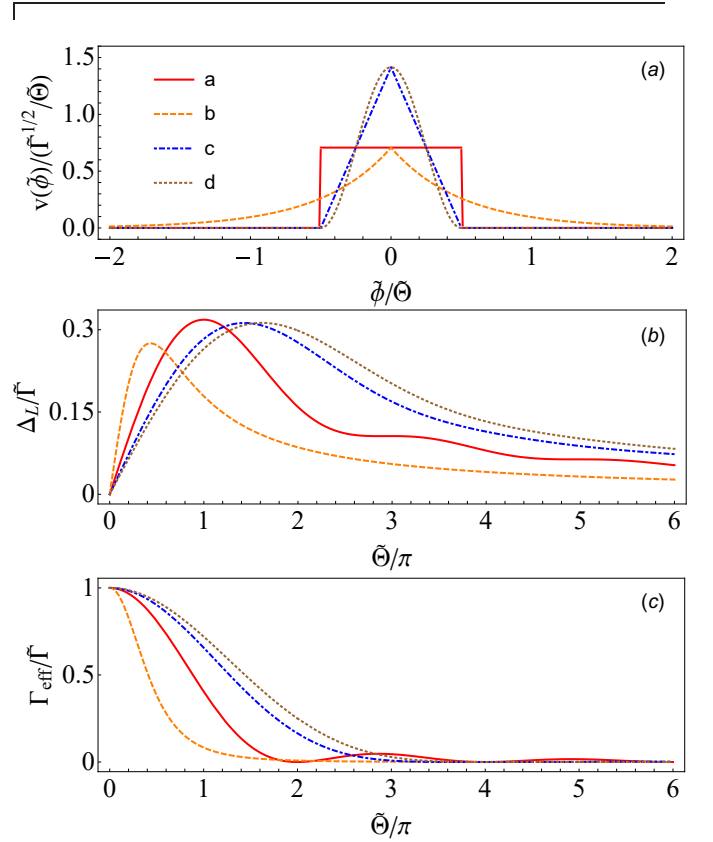}
	\caption{(a) Coupling strength distributions as functions of scaled phase $\tilde{\phi}/\tilde{\Theta}$. 
	(b) and (c) The corresponding Lamb shifts and effective decay rates as functions of characteristic width $\tilde{\Theta}$. 
	Curve a: uniform distribution (red solid line), 
	curve b: exponential distribution (orange dashed line), curve c: triangular distribution (blue dot-dashed line), 
	curve d: raised cosine distribution (brown dotted line).
	}
	\label{continuummarkov}
\end{figure}
In the Markovian limit $T\ll1/\tilde{\Gamma}$, similar to the discrete case, we can replace the phase factor $\phi$ in Eqs.~\eqref{transmissionc}-\eqref{reflectionc} 
by $\tilde{\phi}$. If we define Lamb shift and effective decay for the continuum case as
\begin{subequations}
	\begin{equation}
	\Delta_{\mathrm{L}}=\int\mathrm{d}\tilde{\phi}\mathrm{d}\tilde{\phi}'v\left(\tilde{\phi}\right)v\left(\tilde{\phi}'\right)\sin\left|\tilde{\phi}-\tilde{\phi}'\right|,
	\label{lambshiftc}
	\end{equation}
	\begin{equation}
	\Gamma_{\mathrm{eff}}=2\int\mathrm{d}\tilde{\phi}\mathrm{d}\tilde{\phi}'v\left(\tilde{\phi}\right)v\left(\tilde{\phi}'\right)\cos\left(\tilde{\phi}-\tilde{\phi}'\right),
	\label{gammaeffc}
	\end{equation}
\end{subequations}
the transmission and reflection coefficients can be expressed by Eqs.~\eqref{transmissionMarkov} 
and \eqref{reflectionMarkov}. The scattering spectra exhibit standard Lorentzian lineshapes, 
with $\Delta_{\mathrm{L}}$ and $\Gamma_{\mathrm{eff}}$ [The analytic expressions of them for distributions (a)-(e) are provided 
in Appendix \ref{CouplingDistribution}.] determining all features of the spectra. 
Thus in the following, we concentrate on the influences of coupling distributions on these two quantities. 

First we discuss the case that the coupling distribution is concentrated around one point, as shown in Figs.~\ref{continuumlimit}(a)-\ref{continuumlimit}(d). 
The comparison of these distribution functions with the same characteristic width $\tilde{\Theta}$ is shown in Fig.~\ref{continuummarkov}(a). 
For a given width, the degrees of localization of these distributions follow the following order: $(\mathrm{d})>(\mathrm{c})>(\mathrm{a})>(\mathrm{b})$. 
The corresponding Lamb shifts $\Delta_{\mathrm{L}}$ and effective decay rates $\Gamma_{\mathrm{eff}}$ 
as functions of characteristic width $\tilde{\Theta}$ are shown in Fig.~\ref{continuummarkov}(b) and~\ref{continuummarkov}(c). 
It is not surprising that for all cases, when $\tilde{\Theta}\rightarrow0$, the atom becomes point-like, 
with $\Delta_{\mathrm{L}}\rightarrow0$ and $\Gamma_{\mathrm{eff}}\rightarrow\tilde{\Gamma}$ (i.e., the results under the dipole-approximation). 
For each coupling distribution, with the width $\tilde{\Theta}$ increasing, $\Delta_{\mathrm{L}}$ will first increase and then decrease. 
For a more localized distribution, $\Delta_{\mathrm{L}}$ reaches the maximum slower as $\tilde{\Theta}$ increases [see Fig.~\ref{continuummarkov}(b)]. 
On the other hand, the increase of  $\tilde{\Theta}$ will cause the decrease of $\Gamma_{\mathrm{eff}}$, down to $0$ for large $\tilde{\Theta}$. 
For a more localized distribution, $\Gamma_{\mathrm{eff}}$ decreases slower with the increasing of $\tilde{\Theta}$ [see Fig.~\ref{continuummarkov}(c)]. 
The reason for this is that, when the couplings distribute over an area, due to phase delays between photons re-emitted from different positions, 
the constructive interference like the case of point-like coupling cannot always achieved. 
Particularly, for uniform distribution the atom decouples from the waveguide when $\tilde{\Theta}=2n\pi$ ($n\in\mathbb{N}^{+}$), 
as shown by the solid line in Fig.~\ref{continuummarkov}(c). 
In fact, when $\tilde{\Theta}=2n\pi$, the coupling region can always be divided into many pairs of points with phase delay $\pi$, 
and moreover the re-emitted fields from these points have the same intensitys, resulting in destructive interference and therefore vanished effective decay rate.
\begin{figure}[t]
\includegraphics[width=0.5\textwidth]{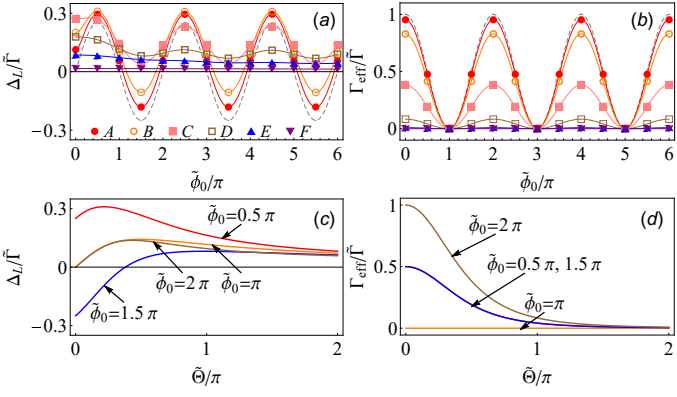}
\caption{Lamb shifts $\Delta_{\mathrm{L}}$ and effective decay rates $\Gamma_{\mathrm{eff}}$ of configuration (e) in 
Fig.~\ref{continuumlimit} for different $\tilde{\phi}_0$ and $\tilde{\Theta}$. 
	(a) and (b) The Lamb shifts $\Delta_{\mathrm{L}}$ and effective decay rates $\Gamma_{\mathrm{eff}}$ as functions of 
	$\tilde{\phi}_0$ with different $\tilde{\Theta}$.
	A: $\tilde{\Theta}=0.1\pi$; B: $\tilde{\Theta}=0.2\pi$; C: $\tilde{\Theta}=0.5\pi$; D: $\tilde{\Theta}=\pi$; 
	E: $\tilde{\Theta}=2\pi$; F: $\tilde{\Theta}=10\pi$; 
	grey dashed lines: $\tilde{\Theta}\to0$, corresponding to the case of two discrete coupling points. 
	(c) and (d) The Lamb shifts $\Delta_{\mathrm{L}}$ and effective decay rates $\Gamma_{\mathrm{eff}}$ 
	as functions of $\tilde{\Theta}$ with different $\tilde{\phi}_0$.}
\label{continuummarkov2}
\end{figure}

Then we discuss the double-exponential distribution shown in Fig.~\ref{continuumlimit}(e), 
where the coupling distribution is concentrated around two points with a phase delay $\tilde{\phi}_0$. 
The characteristic width of the coupling distribution around each point is $\tilde{\Theta}$. 
The Lamb shifts $\Delta_{\mathrm{L}}$ and effective decay rates $\Gamma_{\mathrm{eff}}$ as functions of $\tilde{\phi}_0$ 
with different $\tilde{\Theta}$ are illustrated in Fig.~\ref{continuummarkov2}(a) and~\ref{continuummarkov2}(b), respectively. 
Under the limit $\tilde{\Theta}\to0$, the system reduces to the discrete case with two coupling points, the corresponding Lamb shift and effective decay rate 
are shown by the grey dashed lines in Fig.~\ref{continuummarkov2}(a) and \ref{continuummarkov2}(b) as reference. 
When $\tilde{\Theta}\to0$, the Lamb shift $\Delta_{\mathrm{L}}\to(\tilde{\Gamma}/4)\sin\tilde{\phi}_0$, as shown by the grey dashed line in Fig.~\ref{continuummarkov2}(a). 
As $\tilde{\Theta}$ increases while $\tilde{\Theta}<2\pi$, the curves deviate from the sine function, 
but the oscillation characteristic still remains [see curves A-D in Fig.~\ref{continuummarkov2}(a)]. 
As $\tilde{\Theta}$ continues to increase, this oscillation characteristic disappears [see curves E and F in Fig.~\ref{continuummarkov2}(a)]. 
Under the $\tilde{\Theta}\to0$ limit the effective decay rate $\Gamma_{\mathrm{eff}}\to\tilde{\Gamma}(1+\cos\tilde{\phi}_0)/2$, 
as shown by the grey dashed line in Fig.~\ref{continuummarkov2}(b). As $\tilde{\Theta}$ increases, 
$\Gamma_{\mathrm{eff}}$ still remains a cosine function but with a smaller maximum $16\tilde{\Gamma}/{(\tilde{\Theta}^2+4)^2}$ [see curves A-F in Fig.~\ref{continuummarkov2}(b)], 
due to the coupling-decentralization effect. When $\tilde{\Theta}\geq2\pi$, the effective decay rates are close to $0$ for all $\tilde{\phi}_0$. 
One can also see from these curves that when $\tilde{\phi}_0=(2n-1)\pi$ ($n\in\mathbb{N}^{+}$), 
the atom decouples from the waveguide with $\Gamma_{\mathrm{eff}}=0$ due to destructive interference. In summary, when the 
coupling strength around each coupling point is delocalized, the atom-waveguide coupling as well as the phase-accumulated effects 
become weakened. We also illustrate in Fig.~\ref{continuummarkov2}(c) and \ref{continuummarkov2}(d) the Lamb shifts and effective decay rates as functions of $\tilde{\Theta}$ with different $\tilde{\phi}_0$. With the increase of $\tilde{\Theta}$, the Lamb shifts will first increase and then decrease. 
The effective decay rates for different $\tilde{\phi}_0$ [except for some special values $\tilde{\phi}_0=(2n-1)\pi$, where 
$\Gamma_{\mathrm{eff}}=0$ always holds] decrease to 0 as $\tilde{\Theta}$ increases. 
 \subsection{\label{SpectraContinuumNonMarkovian}Spectra under the continuum limit: the non-Markovian regime}
 \begin{figure}[t]
 	\includegraphics[width=0.45\textwidth]{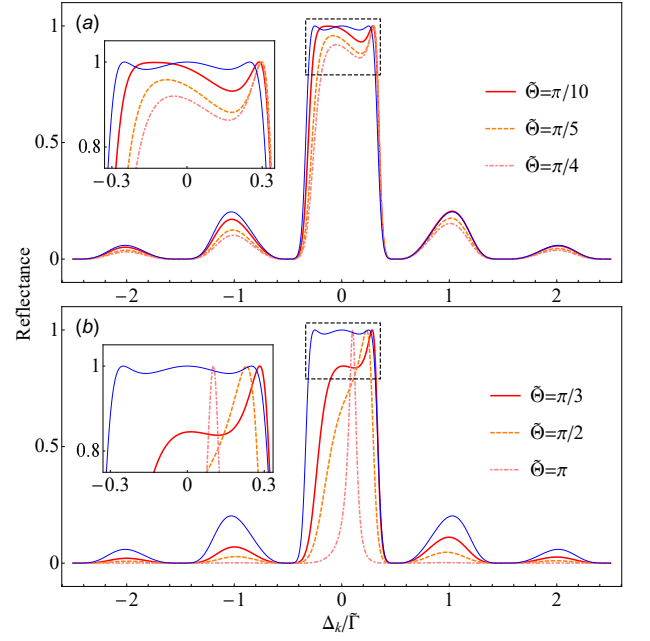}
 	\caption{Reflectances of configuration (e) in Fig.~\ref{continuumlimit} for fixed $\tilde{\phi}_0=500\pi$ and different $\tilde{\Theta}$.
 	         Here $\omega_{\mathrm{a}}=250\tilde{\Gamma}$.
 		(a) $\tilde{\Theta}=\pi/10$ (red solid line), $\tilde{\Theta}=\pi/5$ (orange dashed line), $\tilde{\Theta}=\pi/4$ (pink dot-dashed line). 
 		(b) $\tilde{\Theta}=\pi/3$ (red solid line), $\tilde{\Theta}=\pi/2$ (orange dashed line), $\tilde{\Theta}=\pi$ (pink dot-dashed line). 
		The blue thin lines in (a) and (b) are the reflectance under the limit $\tilde{\Theta}\to0$, where the system reduces to the discrete 
		case with two coupling points.
 		The insets are the reflectances in the neighborhood of $\Delta_{k}$.}
 	\label{continuumnonmarkov}
 \end{figure}
If the coupling distribution between the atom and the waveguide is concentrated around one point [e.g., cases (a)-(d) in Fig.~\ref{continuumlimit}], 
the non-Markovian condition $T>1/\tilde{\Gamma}$ requires $\tilde{\Theta}>\omega_a/\tilde{\Gamma}\gg1$. However, 
one can see from Fig.~\ref {continuummarkov}(b) and \ref {continuummarkov}(c) that $\Delta_{\mathrm{L}}$ and $\Gamma_{\mathrm{eff}}$ vanish even for relatively small $\tilde{\Theta}$. 
Thus it's hard to realize non-Markovianity in these configurations. But for structures with two coupling-concentrated 
regions [e.g., case (e) in Fig.~\ref{continuumlimit}], 
non-Markovianity can also be realized by setting $\tilde{\phi}_0>\omega_a/\tilde{\Gamma}$. The analytic expressions of the 
scattering coefficients for this case are given in Appendix \ref{appendixB}.  
The reflection spectra of the configuration (e) for fixed $\tilde{\phi}_0=500\pi$ (satisfying the non-Markovian condition) 
and different $\tilde{\Theta}$ are shown in Fig.~\ref{continuumnonmarkov}. For comparison, 
we plot the reflection spectra under the limit $\tilde{\Theta}\to0$ [blue thin lines in  Fig.~\ref{continuumnonmarkov}(a) and \ref{continuumnonmarkov}(b)], 
where the system reduces to the discrete case with two coupling points. The remarkable features of the spectra in this parameter region are: 
(1) a symmetric lineshape with three total reflection points in the central area, (2) some side peaks and  total transmission points due to phase-accumulated effects for detuned photons, 
as discussed in Sec.~\ref{SpectraDiscreteNonMarkov}.
With $\tilde{\Theta}$ increases, the reflection spectra become asymmetric, 
while the total transmission points determined by relation $\cos{(1+\Delta_k/\omega_a)}\tilde{\phi}_0+1=0$ still hold. 
In addition, the number of total reflection points changes from three to one [Fig.~\ref{continuumnonmarkov}(a)]. 
As $\tilde{\Theta}$ continues to increase to $\pi$, the width of the central peak decreases and the side peaks gradually disappear [Fig.~\ref{continuumnonmarkov}(b)]. 
In summary, when the coupling strength around each coupling point is delocalized, 
the phase-accumulated effects due to detuning become weakened. 
\section{\label{conclusion}CONCLUSIONS AND DISCUSSIONS}
We have investigated the single-photon scattering spectra of a single two-level giant atom coupled to a one dimensional waveguide via 
$N$ connection points or a continuous coupling region. We provide the general analytical solutions for the single-photon transmission and 
reflection coefficients, which are valid in both the Marlovian and the non-Marlovian regimes. 

For the case of discrete coupling points,
we analyze the features of the scattering spectra in the Markovian, the moderately non-Markovian and the deep non-Markovian regimes, 
respectively. We find that in the Markovian regime, the influence of detuning on the phase decay can be ignored, resulting in Lorentzian
lineshapes characterized by the Lamb shifts and the effective decay rates. While in the non-Markovian regime, the accumulated phases 
become detuning-dependent, giving rise to non-Lorentzian lineshapes, characterized by multiple side peaks and total transmission 
points. Anther remarkable and interesting feature in this regime is appearance of broad photonic band gap when the parameters are 
appropriately designed.  We further obtain the results for the continuum limit with the atom couples to the waveguide through a continuous region.
We analyze and compare the scattering spectra for some typical distributions of coupling strength, and summarize the influences of non-dipole effects on the scattering spectra.

Our results show that the single-photon scattering spectra could be powerful tools to probe non-dipole effects and light-matter interactions in giant-atom waveguide-QED structures. Additionally, based on the analytical solutions for the discrete and the continuum case, it is possible to control photon transport on demand by suitably design the coupling configuration.

Finally, we wish to make some remark on the difference between our model and waveguide-QED systems with multiple small atoms 
\cite{Chang-NJP2012, Greenberg-PRA2021, Mukhopadhyay-PRA2020,Poudyal-PRR2020}. For a waveguide-QED system 
with multiple small atoms, due to photon exchange between different atoms, photon mediated interactions between atoms 
and collective decays appear. In addition, a single photon can excite an atomic chain to its collective excited states. Accordingly, the scattering spectra can exhibit some feathers like those in the atomic coherence phenomena \cite{Mukhopadhyay-PRA2020,Poudyal-PRR2020}. While for a single giant atom there are not atom-atom interactions, and photons can only excite single-atom states instead of collective states. But in this situation, there exists photon exchange between different connecting points and interfere effects between decayed photons from these points. Thus the atom will obtain phase-dependent frequency shift  and effective decay. The scattering spectra will be strongly influenced by these quantities.  
\begin{acknowledgments}
This work was supported by the National Natural Science Foundation of China (NSFC) under Grants No. 
11404269, No. 61871333, and No. 12047576.
\end{acknowledgments}
\appendix
\section{\label{ContinuumHamiltonian} Hamiltonian and single-photon wave function under the continuum limit}
According to the Hamiltonian \eqref{totalHamiltonian} of the discrete case, the Hamiltonian under the continuum limit can be derived as
\begin{eqnarray}
\hat{H}&=&
\omega_\mathrm{a}\left|e\right>\left<e\right|+
\int\mathrm{d}x\hat{c}_\mathrm{R}^{\dagger}\left(x\right)\left(-iv_{\mathrm{g}}\frac{\partial}{\partial x}\right)\hat{c}_\mathrm{R}\left(x\right) 
\nonumber \\
&&+\int\mathrm{d}x\hat{c}_\mathrm{L}^{\dagger}\left(x\right)\left(iv_{\mathrm{g}}\frac{\partial}{\partial x}\right)\hat{c}_\mathrm{L}\left(x\right)
\nonumber \\
&&+\int\mathrm{d}x V(x)\Bigg[\sum_{\mathrm{i}=\mathrm{R},\mathrm{L}}\hat{c}_\mathrm{i}^{\dagger}\left(x\right)\sigma_{-}+\mathrm{H}.\mathrm{c}.\Bigg],
\label{Hamiltonianofcontinuum}
\end{eqnarray}
where $V(x)=\omega_{\mathrm{a}}v(\tilde{\phi}) /\sqrt{v_{\mathrm{g}}} $. $v(\tilde{\phi})$ is the distribution of coupling strength defined in Sec.~\ref{Continuum limit}, satisfying $\int\mathrm{d}\tilde{\phi}v\left(\tilde{\phi}\right)=(\tilde{\Gamma}/2)^{1/2}$, with $\tilde{\phi}=\omega_\mathrm{a}x/v_\mathrm{g}$. According to Eqs.~\eqref{eigenstate} and \eqref{eigenequation}, we can obtain 
\begin{subequations}
	\begin{equation}
	\left(-iv_\mathrm{g}\frac{\partial}{\partial x}-E\right)\Phi_\mathrm{R}\left(x\right)+V(x)f_\mathrm{a}=0,
	\label{EQMC1}
	\end{equation}
	\begin{equation}
	\left(iv_\mathrm{g}\frac{\partial}{\partial x}-E\right)\Phi_\mathrm{L}\left(x\right)+V(x)f_\mathrm{a}=0,
	\label{EQMC2}
	\end{equation}
	\begin{equation}
	\left(\omega_\mathrm{a}-E\right)f_\mathrm{a}+ \int\mathrm{d}x V(x)\left[\Phi_\mathrm{R}\left(x\right)+\Phi_\mathrm{L}\left(x\right)\right]=0.
	\label{EQMC3}
	\end{equation}
\end{subequations}

The wave functions of a single photon initially coming from left will be
\begin{subequations}
	\begin{eqnarray}
	\Phi_\mathrm{R}\left(x\right)=e^{ikx}t_\mathrm{c}\left(x\right),
	\label{CWF1}
	\end{eqnarray}
	\begin{equation}
	\Phi_\mathrm{L}\left(x\right)=e^{-ikx}r_\mathrm{c}\left(x\right),
	\label{CWF2}
	\end{equation}
\end{subequations}
where $t_\mathrm{c}(x)$ [$r_\mathrm{c}(x)$] is the probability amplitude for a right-going (left-going) photon appearing at position $x$,
satisfying the boundary condition $t_\mathrm{c}(-\infty)=1$ [$r_\mathrm{c}(+\infty)=0$]. And the scattering amplitudes to be fixed can be defined as
 $t=t_\mathrm{c}(+\infty)$ and $r=r_\mathrm{c}(-\infty)$.
 Substituting Eqs.~\eqref{CWF1} and \eqref{CWF2} into Eqs.~\eqref{EQMC1} and \eqref{EQMC2}, one can obtain the following relations
 \begin{subequations}
 	\begin{eqnarray}
t_\mathrm{c}\left(x\right)=1-i\frac{f_{\mathrm{a}}}{v_\mathrm{g}}\int_{-\infty}^{x}V\left(x\right)e^{-ikx}\mathrm{d}x,
 	\label{output1}
 	\end{eqnarray}
 	\begin{equation}
r_\mathrm{c}\left(x\right)=-i\frac{f_{\mathrm{a}}}{v_\mathrm{g}}\int_{x}^{\infty}V\left(x\right)e^{ikx}\mathrm{d}x.
 	\label{output2}
 	\end{equation}
 \end{subequations}
Plugging these relations into  Eq.~\eqref{EQMC3}, one can solve for the excitation amplitude of the atom
	\begin{equation}
f_{\mathrm{a}}=\frac{i\sqrt{v_{\mathrm{g}}}\int\mathrm{d}\tilde{\phi}v\left(\tilde{\phi}\right)e^{i\phi}}{i\Delta_{k}-\int\mathrm{d}\tilde{\phi}\mathrm{d}\tilde{\phi}'v\left(\tilde{\phi}\right)v\left(\tilde{\phi}'\right)e^{i\left|\phi-\phi'\right|}},
\label{excitation}
	\end{equation}
where  $\phi=kx=(1+\Delta_{k}/\omega_\mathrm{a})\tilde{\phi}$. Substituting Eq.~\eqref{excitation} into Eqs.~\eqref{output1} and \eqref{output2}, one can obtain the scattering amplitudes
 \begin{subequations}
	\begin{eqnarray}
t=t_{\mathrm{c}}\left(+\infty\right)=\frac{{i\Delta_{k}-i\int\mathrm{d}\tilde{\phi}\mathrm{d}\tilde{\phi}'v\left(\tilde{\phi}\right)v\left(\tilde{\phi}'\right)\sin\left|\phi-\phi'\right|}}{i\Delta_{k}-\int\mathrm{d}\tilde{\phi}\mathrm{d}\tilde{\phi}'v\left(\tilde{\phi}\right)v\left(\tilde{\phi}'\right)e^{i\left|\phi-\phi'\right|}},
	\nonumber\\
	\label{continuumt}
	\end{eqnarray}
	\begin{equation}
	r=r_{\mathrm{c}}\left(-\infty\right)=\frac{\left[\int\mathrm{d}\tilde{\phi}\mathrm{d}\tilde{\phi}'v\left(\tilde{\phi}\right)v\left(\tilde{\phi}'\right)\cos\left(\phi-\phi'\right)\right]e^{i\alpha}}{i\Delta_{k}-\int\mathrm{d}\tilde{\phi}\mathrm{d}\tilde{\phi}'v\left(\tilde{\phi}\right)v\left(\tilde{\phi}'\right)e^{i\left|\phi-\phi'\right|}},
	\label{continuumr}
	\end{equation}
\end{subequations}
where 
	\begin{equation}
	\tan\alpha=\frac{\int\mathrm{d}\tilde{\phi}\mathrm{d}\tilde{\phi}'v\left(\tilde{\phi}\right)v\left(\tilde{\phi}'\right)\sin\left(\phi+\phi'\right)}{\int\mathrm{d}\tilde{\phi}\mathrm{d}\tilde{\phi}'v\left(\tilde{\phi}\right)v\left(\tilde{\phi}'\right)\cos\left(\phi+\phi'\right)}.
	\nonumber
\end{equation}
Note that in above derivations, the relation $\int\mathrm{d}xV(x)=\sqrt{v_{\mathrm{g}}}\int\mathrm{d}\tilde{\phi}v(\tilde{\phi})$ is used.
Finally, by using the definitions $\mathcal{T}=|t|^2$ and $\mathcal{R}=|r|^2$, one can arrive at Eqs.~\eqref{transmissionc} and \eqref{reflectionc} in the main text.
\section{\label{CouplingDistribution} Expressions of coupling-distribution functions, Lamb shifts and effective decays}
In Fig.~\ref{continuumlimit}, we have schematically shown five different coupling distributions: (a) uniform distribution, (b) exponential distribution, (c) triangular distribution, (d) raised cosine distribution, (e) double exponential distribution. The corresponding expressions for these distributions are summarized as follows:
\begin{subequations}
	\begin{equation}
	v_\mathrm{a}\left(\tilde{\phi}\right)=\left(\frac{\tilde{\Gamma}}{2}\right)^{\frac{1}{2}}\frac{1}{\tilde{\Theta}},~~~{-\frac{\tilde{\Theta}}{2}\leq\tilde{\phi}\leq\frac{\tilde{\Theta}}{2}}
	\label{va}
	\end{equation}
	\begin{equation}
	v_\mathrm{b}\left(\tilde{\phi}\right)=\left(\frac{\tilde{\Gamma}}{2}\right)^{\frac{1}{2}}\frac{1}{\tilde{\Theta}}e^{-\frac{2\left|\tilde{\phi}\right|}{\tilde{\Theta}}},
	\label{vb}
	\end{equation}
	\begin{equation}
	v_\mathrm{c}\left(\tilde{\phi}\right)= \left(\frac{\tilde{\Gamma}}{2}\right)^{\frac{1}{2}}\frac{2}{\tilde{\Theta}}\left(1-\frac{2\left|\tilde{\phi}\right|}{\tilde{\Theta}}\right),~~~{-\frac{\tilde{\Theta}}{2}\leq\tilde{\phi}\leq\frac{\tilde{\Theta}}{2}}
	\label{vc}
	\end{equation}
	\begin{equation}
	v_\mathrm{d}\left(\tilde{\phi}\right)= \left(\frac{\tilde{\Gamma}}{2}\right)^{\frac{1}{2}}\frac{2}{\tilde{\Theta}}\cos^2\left(\frac{\tilde{\phi}}{\tilde{\Theta}}\pi\right),~~~{-\frac{\tilde{\Theta}}{2}\leq\tilde{\phi}\leq\frac{\tilde{\Theta}}{2}}
	\label{vd}
	\end{equation}
	\begin{equation}
	v_\mathrm{e}\left(\tilde{\phi}\right)=\left(\frac{\tilde{\Gamma}}{2}\right)^{\frac{1}{2}}\frac{1}{2\tilde{\Theta}}\left(e^{-\frac{2}{\tilde{\Theta}}\left|\tilde{\phi}+\frac{\tilde{\phi}_0}{2}\right|}+e^{-\frac{2}{\tilde{\Theta}}\left|\tilde{\phi}-\frac{\tilde{\phi}_0}{2}\right|}\right).
	\label{ve}
	\end{equation}
\end{subequations}
 One can check that $\int v_i\left(\tilde{\phi}\right)d\tilde{\phi}=(\tilde{\Gamma}/2)^{1/2}$ ($i=\mathrm{a},~\mathrm{b},~\mathrm{c},~\mathrm{d},~\mathrm{e}$) for
 each distribution, where $\tilde{\Gamma}$ is the decay rate under the dipole-approximation limit.
 By substituting the distribution functions Eqs.~\eqref{va}-\eqref{ve} into Eq.~\eqref{lambshiftc}, one can obtain the corresponding analytic expressions of the Lamb shifts
\begin{subequations}
	\begin{equation}
	\Delta_{\mathrm{L}}^{\left(\mathrm{a}\right)}=\tilde{\Gamma}\frac{\tilde{\Theta}-\sin\tilde{\Theta}}{\tilde{\Theta}^2},
	\label{Lambshifta}
	\end{equation}
	\begin{equation}
	\Delta_{\mathrm{L}}^{\left(\mathrm{b}\right)}=\tilde{\Gamma}\frac{\tilde{\Theta}\left(\tilde{\Theta}^2+12\right)}{2\left(\tilde{\Theta}^2+4\right)^2},
	\label{lambshiftb}
	\end{equation}
	\begin{equation}
	\Delta_{\mathrm{L}}^{\left(\mathrm{c}\right)}=\frac{4\tilde{\Gamma}} {3\tilde{\Theta}^4}{\left(12\tilde{\Theta}+\tilde{\Theta}^3-48\sin\frac{\tilde{\Theta}}{2}+12\sin\tilde{\Theta}\right)},
	\label{Lambshiftc}
	\end{equation}
	\begin{equation}
	\Delta_{\mathrm{L}}^{\left(\mathrm{d}\right)}=\tilde{\Gamma}\frac{32\pi^4\tilde{\Theta}-20\pi^2\tilde{\Theta}^3+3\tilde{\Theta}^5-32\pi^4\sin\tilde{\Theta}}
	{2\left(\tilde{\Theta}^3-4\pi^2\tilde{\Theta}\right)^2},
	\label{Lambshiftd}
	\end{equation}
	\begin{eqnarray}
	\Delta_{\mathrm{L}}^{\left(\mathrm{e}\right)}&=&\frac{\tilde{\Gamma}}{4\left(\tilde{\Theta}^2+4\right)^2}
	\bigg[\left(8\tilde{\phi_0}+12\tilde{\Theta}+\tilde{\Theta}^3+2\tilde{\Theta}^2\tilde{\phi_0}\right)e^{-\frac{2\tilde{\phi}_0}{\tilde{\Theta}}}
	\nonumber \\
	&&+12\tilde{\Theta}+\tilde{\Theta}^3+16\sin\tilde{\phi}_0\bigg].
	\label{Lambshifte}
	\end{eqnarray}
	\end{subequations}
And by substituting Eqs.~\eqref{va}-~\eqref{ve} into Eq.~\eqref{gammaeffc}, one can get the corresponding effective decay rates 
\begin{subequations}
	\begin{equation}
	\Gamma_{\mathrm{eff}}^{\left(\mathrm{a}\right)}=2\tilde{\Gamma}\frac{1-\cos\tilde{\Theta}}{\tilde{\Theta}^2},
	\label{Gammaeffa}
	\end{equation}
	
	\begin{equation}
	\Gamma_{\mathrm{eff}}^{\left(\mathrm{b}\right)}=\tilde{\Gamma}\frac{16}{\left(\tilde{\Theta}^2+4\right)^2},
	\label{Gammaeffb}
	\end{equation}
	\begin{equation}
	\Gamma_{\mathrm{eff}}^{\left(\mathrm{c}\right)}=\tilde{\Gamma}\frac{256\sin^4\left(\frac{\tilde{\Theta}}{4}\right)}
	{\tilde{\Theta}^4},
	\label{Gammaeffc}
	\end{equation}
	\begin{equation}
	\Gamma_{\mathrm{eff}}^{\left(\mathrm{d}\right)}=\tilde{\Gamma}\frac{32\pi^4\left(1-\cos\tilde{\Theta}\right)}
	{\left(\tilde{\Theta}^3-4\pi^2\tilde{\Theta}\right)^2},
	\label{Gammaeffd}
	\end{equation}
	\begin{eqnarray}
	\Gamma_{\mathrm{eff}}^{\left(\mathrm{e}\right)}=\tilde{\Gamma}\frac{8\left(1+\cos\tilde{\phi}_0\right)}
	{\left(\tilde{\Theta}^2+4\right)^2}.
	\label{Gammaeffe}
	\end{eqnarray}
\end{subequations}
\section{\label{appendixB} Scattering coefficients for the double-exponential distribution}
In this Appendix, we will provide the analytic expressions of the scattering coefficients in the non-Markovian regime for the distribution given by \eqref{ve}.
Substituting Eq.~\ref{ve} into Eqs.~\ref{transmissionc} and~\ref{reflectionc}, one can obtain the transmission and reflection coefficients as
\begin{subequations}
	\begin{equation}
	\mathcal{T}=\frac{\left[\Delta_{k}-A\left(\Delta_{k}\right)\right]^2}{\left[\Delta_{k}-A\left(\Delta_{k}\right)\right]^2+\left[B\left(\Delta_{k}\right)\right]^2},
	\label{transmissione}
	\end{equation}
	\begin{equation}
	\mathcal{R}=\frac{\left[B\left(\Delta_{k}\right)\right]^2}{\left[\Delta_{k}-A\left(\Delta_{k}\right)\right]^2+\left[B\left(\Delta_{k}\right)\right]^2},
	\label{reflectione}
	\end{equation}
\end{subequations}
where
\begin{subequations}
	\begin{eqnarray}
	A\left(\Delta_{k}\right)&=&\frac{\Gamma}{4\left(\Theta^2+4\right)^2}
	\bigg[\left(8\phi_0+12\Theta+\Theta^3+2\Theta^2\phi_0\right)e^{-\frac{2\tilde{\phi}_0}{\tilde{\Theta}}}
	\nonumber \\
	&&+12\Theta+\Theta^3+16\sin\phi_0\bigg],
	\label{A}
	\end{eqnarray}
	\begin{equation}
	B\left(\Delta_{k}\right) =\tilde{\Gamma}\frac{4\left(1+\cos\phi_0\right)}{\left(\Theta^2+4\right)^2},
	\end{equation}
\end{subequations}
with $\Theta$ and $\phi_0$ being defined as $\Theta=(1+\Delta_{k}/\omega_\mathrm{a})\tilde{\Theta}$ and  $\phi_0 =(1+\Delta_{k}/\omega_\mathrm{a})\tilde{\phi_0}$,
respectively. 

\bibliography{qy-Aug-28-2021}

\providecommand{\noopsort}[1]{}\providecommand{\singleletter}[1]{#1}%
\begin{thebibliography}{37}%
\makeatletter
\providecommand \@ifxundefined [1]{%
 \@ifx{#1\undefined}
}%
\providecommand \@ifnum [1]{%
 \ifnum #1\expandafter \@firstoftwo
 \else \expandafter \@secondoftwo
 \fi
}%
\providecommand \@ifx [1]{%
 \ifx #1\expandafter \@firstoftwo
 \else \expandafter \@secondoftwo
 \fi
}%
\providecommand \natexlab [1]{#1}%
\providecommand \enquote  [1]{``#1''}%
\providecommand \bibnamefont  [1]{#1}%
\providecommand \bibfnamefont [1]{#1}%
\providecommand \citenamefont [1]{#1}%
\providecommand \href@noop [0]{\@secondoftwo}%
\providecommand \href [0]{\begingroup \@sanitize@url \@href}%
\providecommand \@href[1]{\@@startlink{#1}\@@href}%
\providecommand \@@href[1]{\endgroup#1\@@endlink}%
\providecommand \@sanitize@url [0]{\catcode `\\12\catcode `\$12\catcode
  `\&12\catcode `\#12\catcode `\^12\catcode `\_12\catcode `\%12\relax}%
\providecommand \@@startlink[1]{}%
\providecommand \@@endlink[0]{}%
\providecommand \url  [0]{\begingroup\@sanitize@url \@url }%
\providecommand \@url [1]{\endgroup\@href {#1}{\urlprefix }}%
\providecommand \urlprefix  [0]{URL }%
\providecommand \Eprint [0]{\href }%
\providecommand \doibase [0]{http://dx.doi.org/}%
\providecommand \selectlanguage [0]{\@gobble}%
\providecommand \bibinfo  [0]{\@secondoftwo}%
\providecommand \bibfield  [0]{\@secondoftwo}%
\providecommand \translation [1]{[#1]}%
\providecommand \BibitemOpen [0]{}%
\providecommand \bibitemStop [0]{}%
\providecommand \bibitemNoStop [0]{.\EOS\space}%
\providecommand \EOS [0]{\spacefactor3000\relax}%
\providecommand \BibitemShut  [1]{\csname bibitem#1\endcsname}%
\let\auto@bib@innerbib\@empty
\bibitem [{\citenamefont {Walls}\ and\ \citenamefont
  {Milburn}(2008)}]{Walls-QuantumOptics}%
  \BibitemOpen
  \bibfield  {author} {\bibinfo {author} {\bibfnamefont {D.}~\bibnamefont
  {Walls}}\ and\ \bibinfo {author} {\bibfnamefont {G.~J.}\ \bibnamefont
  {Milburn}},\ }\href@noop {} {\emph {\bibinfo {title} {Quantum Optics}}}\
  (\bibinfo  {publisher} {Springer-Verlag Berlin Heidelberg},\ \bibinfo {year}
  {2008})\BibitemShut {NoStop}%
\bibitem [{\citenamefont {Roy}\ \emph {et~al.}(2017)\citenamefont {Roy},
  \citenamefont {Wilson},\ and\ \citenamefont {Firstenberg}}]{Roy-PRM2017}%
  \BibitemOpen
  \bibfield  {author} {\bibinfo {author} {\bibfnamefont {D.}~\bibnamefont
  {Roy}}, \bibinfo {author} {\bibfnamefont {C.~M.}\ \bibnamefont {Wilson}}, \
  and\ \bibinfo {author} {\bibfnamefont {O.}~\bibnamefont {Firstenberg}},\
  }\href {\doibase 10.1103/RevModPhys.89.021001} {\bibfield  {journal}
  {\bibinfo  {journal} {Rev. Mod. Phys.}\ }\textbf {\bibinfo {volume} {89}},\
  \bibinfo {pages} {021001} (\bibinfo {year} {2017})}\BibitemShut {NoStop}%
\bibitem [{\citenamefont {Gu}\ \emph {et~al.}(2017)\citenamefont {Gu},
  \citenamefont {Kockum}, \citenamefont {Miranowicz}, \citenamefont {Liu},\
  and\ \citenamefont {Nori}}]{Gu-PhysReports2017}%
  \BibitemOpen
  \bibfield  {author} {\bibinfo {author} {\bibfnamefont {X.}~\bibnamefont
  {Gu}}, \bibinfo {author} {\bibfnamefont {A.~F.}\ \bibnamefont {Kockum}},
  \bibinfo {author} {\bibfnamefont {A.}~\bibnamefont {Miranowicz}}, \bibinfo
  {author} {\bibfnamefont {Y.-x.}\ \bibnamefont {Liu}}, \ and\ \bibinfo
  {author} {\bibfnamefont {F.}~\bibnamefont {Nori}},\ }\href {\doibase
  https://doi.org/10.1016/j.physrep.2017.10.002} {\bibfield  {journal}
  {\bibinfo  {journal} {Phys. Rep.}\ }\textbf {\bibinfo {volume} {718-719}},\
  \bibinfo {pages} {1 } (\bibinfo {year} {2017})}\BibitemShut {NoStop}%
\bibitem [{\citenamefont {Gustafsson}\ \emph {et~al.}(2014)\citenamefont
  {Gustafsson}, \citenamefont {Aref}, \citenamefont {Kockum}, \citenamefont
  {Ekstr{\"o}m}, \citenamefont {Johansson},\ and\ \citenamefont
  {Delsing}}]{Gustafsson-Science2014}%
  \BibitemOpen
  \bibfield  {author} {\bibinfo {author} {\bibfnamefont {M.~V.}\ \bibnamefont
  {Gustafsson}}, \bibinfo {author} {\bibfnamefont {T.}~\bibnamefont {Aref}},
  \bibinfo {author} {\bibfnamefont {A.~F.}\ \bibnamefont {Kockum}}, \bibinfo
  {author} {\bibfnamefont {M.~K.}\ \bibnamefont {Ekstr{\"o}m}}, \bibinfo
  {author} {\bibfnamefont {G.}~\bibnamefont {Johansson}}, \ and\ \bibinfo
  {author} {\bibfnamefont {P.}~\bibnamefont {Delsing}},\ }\href {\doibase
  10.1126/science.1257219} {\bibfield  {journal} {\bibinfo  {journal}
  {Science}\ }\textbf {\bibinfo {volume} {346}},\ \bibinfo {pages} {207}
  (\bibinfo {year} {2014})}\BibitemShut {NoStop}%
\bibitem [{\citenamefont {Koch}\ \emph {et~al.}(2007)\citenamefont {Koch},
  \citenamefont {Yu}, \citenamefont {Gambetta}, \citenamefont {Houck},
  \citenamefont {Schuster}, \citenamefont {Majer}, \citenamefont {Blais},
  \citenamefont {Devoret}, \citenamefont {Girvin},\ and\ \citenamefont
  {Schoelkopf}}]{Koch-PRA2007}%
  \BibitemOpen
  \bibfield  {author} {\bibinfo {author} {\bibfnamefont {J.}~\bibnamefont
  {Koch}}, \bibinfo {author} {\bibfnamefont {T.~M.}\ \bibnamefont {Yu}},
  \bibinfo {author} {\bibfnamefont {J.}~\bibnamefont {Gambetta}}, \bibinfo
  {author} {\bibfnamefont {A.~A.}\ \bibnamefont {Houck}}, \bibinfo {author}
  {\bibfnamefont {D.~I.}\ \bibnamefont {Schuster}}, \bibinfo {author}
  {\bibfnamefont {J.}~\bibnamefont {Majer}}, \bibinfo {author} {\bibfnamefont
  {A.}~\bibnamefont {Blais}}, \bibinfo {author} {\bibfnamefont {M.~H.}\
  \bibnamefont {Devoret}}, \bibinfo {author} {\bibfnamefont {S.~M.}\
  \bibnamefont {Girvin}}, \ and\ \bibinfo {author} {\bibfnamefont {R.~J.}\
  \bibnamefont {Schoelkopf}},\ }\href {\doibase 10.1103/PhysRevA.76.042319}
  {\bibfield  {journal} {\bibinfo  {journal} {Phys. Rev. A}\ }\textbf {\bibinfo
  {volume} {76}},\ \bibinfo {pages} {042319} (\bibinfo {year}
  {2007})}\BibitemShut {NoStop}%
\bibitem [{\citenamefont {Frisk~Kockum}(2021)}]{Kockum-Arxiv2020}%
  \BibitemOpen
  \bibfield  {author} {\bibinfo {author} {\bibfnamefont {A.}~\bibnamefont
  {Frisk~Kockum}},\ }in\ \href@noop {} {\emph {\bibinfo {booktitle}
  {International Symposium on Mathematics, Quantum Theory, and
  Cryptography}}},\ \bibinfo {editor} {edited by\ \bibinfo {editor}
  {\bibfnamefont {T.}~\bibnamefont {Takagi}}, \bibinfo {editor} {\bibfnamefont
  {M.}~\bibnamefont {Wakayama}}, \bibinfo {editor} {\bibfnamefont
  {K.}~\bibnamefont {Tanaka}}, \bibinfo {editor} {\bibfnamefont
  {N.}~\bibnamefont {Kunihiro}}, \bibinfo {editor} {\bibfnamefont
  {K.}~\bibnamefont {Kimoto}}, \ and\ \bibinfo {editor} {\bibfnamefont
  {Y.}~\bibnamefont {Ikematsu}}}\ (\bibinfo  {publisher} {Springer Singapore},\
  \bibinfo {address} {Singapore},\ \bibinfo {year} {2021})\ pp.\ \bibinfo
  {pages} {125--146}\BibitemShut {NoStop}%
\bibitem [{\citenamefont {Andersson}\ \emph {et~al.}(2019)\citenamefont
  {Andersson}, \citenamefont {Suri}, \citenamefont {Guo}, \citenamefont
  {Aref},\ and\ \citenamefont {Delsing}}]{Andersson-Nature2019}%
  \BibitemOpen
  \bibfield  {author} {\bibinfo {author} {\bibfnamefont {G.}~\bibnamefont
  {Andersson}}, \bibinfo {author} {\bibfnamefont {B.}~\bibnamefont {Suri}},
  \bibinfo {author} {\bibfnamefont {L.}~\bibnamefont {Guo}}, \bibinfo {author}
  {\bibfnamefont {T.}~\bibnamefont {Aref}}, \ and\ \bibinfo {author}
  {\bibfnamefont {P.}~\bibnamefont {Delsing}},\ }\href {\doibase
  10.1038/s41567-019-0605-6} {\bibfield  {journal} {\bibinfo  {journal} {Nature
  Physics}\ }\textbf {\bibinfo {volume} {15}},\ \bibinfo {pages} {1123}
  (\bibinfo {year} {2019})}\BibitemShut {NoStop}%
\bibitem [{\citenamefont {Kannan}\ \emph {et~al.}(2020)\citenamefont {Kannan},
  \citenamefont {Ruckriegel}, \citenamefont {Campbell}, \citenamefont {Kockum},
  \citenamefont {Braum{\"u}ller}, \citenamefont {Kim}, \citenamefont
  {Kjaergaard}, \citenamefont {Krantz}, \citenamefont {Melville}, \citenamefont
  {Niedzielski}, \citenamefont {Veps{\"a}l{\"a}inen}, \citenamefont {Winik},
  \citenamefont {Yoder}, \citenamefont {Nori}, \citenamefont {Orlando},
  \citenamefont {Gustavsson},\ and\ \citenamefont
  {Oliver}}]{Kannan-Nature2020}%
  \BibitemOpen
  \bibfield  {author} {\bibinfo {author} {\bibfnamefont {B.}~\bibnamefont
  {Kannan}}, \bibinfo {author} {\bibfnamefont {M.~J.}\ \bibnamefont
  {Ruckriegel}}, \bibinfo {author} {\bibfnamefont {D.~L.}\ \bibnamefont
  {Campbell}}, \bibinfo {author} {\bibfnamefont {A.~F.}\ \bibnamefont
  {Kockum}}, \bibinfo {author} {\bibfnamefont {J.}~\bibnamefont
  {Braum{\"u}ller}}, \bibinfo {author} {\bibfnamefont {D.~K.}\ \bibnamefont
  {Kim}}, \bibinfo {author} {\bibfnamefont {M.}~\bibnamefont {Kjaergaard}},
  \bibinfo {author} {\bibfnamefont {P.}~\bibnamefont {Krantz}}, \bibinfo
  {author} {\bibfnamefont {A.}~\bibnamefont {Melville}}, \bibinfo {author}
  {\bibfnamefont {B.~M.}\ \bibnamefont {Niedzielski}}, \bibinfo {author}
  {\bibfnamefont {A.}~\bibnamefont {Veps{\"a}l{\"a}inen}}, \bibinfo {author}
  {\bibfnamefont {R.}~\bibnamefont {Winik}}, \bibinfo {author} {\bibfnamefont
  {J.~L.}\ \bibnamefont {Yoder}}, \bibinfo {author} {\bibfnamefont
  {F.}~\bibnamefont {Nori}}, \bibinfo {author} {\bibfnamefont {T.~P.}\
  \bibnamefont {Orlando}}, \bibinfo {author} {\bibfnamefont {S.}~\bibnamefont
  {Gustavsson}}, \ and\ \bibinfo {author} {\bibfnamefont {W.~D.}\ \bibnamefont
  {Oliver}},\ }\href {\doibase 10.1038/s41586-020-2529-9} {\bibfield  {journal}
  {\bibinfo  {journal} {Nature}\ }\textbf {\bibinfo {volume} {583}},\ \bibinfo
  {pages} {775} (\bibinfo {year} {2020})}\BibitemShut {NoStop}%
\bibitem [{\citenamefont {Kockum}\ \emph {et~al.}(2014)\citenamefont {Kockum},
  \citenamefont {Delsing},\ and\ \citenamefont {Johansson}}]{Kockum-PRA2014}%
  \BibitemOpen
  \bibfield  {author} {\bibinfo {author} {\bibfnamefont {A.~F.}\ \bibnamefont
  {Kockum}}, \bibinfo {author} {\bibfnamefont {P.}~\bibnamefont {Delsing}}, \
  and\ \bibinfo {author} {\bibfnamefont {G.}~\bibnamefont {Johansson}},\ }\href
  {\doibase 10.1103/PhysRevA.90.013837} {\bibfield  {journal} {\bibinfo
  {journal} {Phys. Rev. A}\ }\textbf {\bibinfo {volume} {90}},\ \bibinfo
  {pages} {013837} (\bibinfo {year} {2014})}\BibitemShut {NoStop}%
\bibitem [{\citenamefont {Vadiraj}\ \emph {et~al.}(2021)\citenamefont
  {Vadiraj}, \citenamefont {Ask}, \citenamefont {McConkey}, \citenamefont
  {Nsanzineza}, \citenamefont {Chang}, \citenamefont {Kockum},\ and\
  \citenamefont {Wilson}}]{Vadiraj-PRA2021}%
  \BibitemOpen
  \bibfield  {author} {\bibinfo {author} {\bibfnamefont {A.~M.}\ \bibnamefont
  {Vadiraj}}, \bibinfo {author} {\bibfnamefont {A.}~\bibnamefont {Ask}},
  \bibinfo {author} {\bibfnamefont {T.~G.}\ \bibnamefont {McConkey}}, \bibinfo
  {author} {\bibfnamefont {I.}~\bibnamefont {Nsanzineza}}, \bibinfo {author}
  {\bibfnamefont {C.~W.~S.}\ \bibnamefont {Chang}}, \bibinfo {author}
  {\bibfnamefont {A.~F.}\ \bibnamefont {Kockum}}, \ and\ \bibinfo {author}
  {\bibfnamefont {C.~M.}\ \bibnamefont {Wilson}},\ }\href {\doibase
  10.1103/PhysRevA.103.023710} {\bibfield  {journal} {\bibinfo  {journal}
  {Phys. Rev. A}\ }\textbf {\bibinfo {volume} {103}},\ \bibinfo {pages}
  {023710} (\bibinfo {year} {2021})}\BibitemShut {NoStop}%
\bibitem [{\citenamefont {Kockum}\ \emph {et~al.}(2018)\citenamefont {Kockum},
  \citenamefont {Johansson},\ and\ \citenamefont {Nori}}]{Kockum-PRL2018}%
  \BibitemOpen
  \bibfield  {author} {\bibinfo {author} {\bibfnamefont {A.~F.}\ \bibnamefont
  {Kockum}}, \bibinfo {author} {\bibfnamefont {G.}~\bibnamefont {Johansson}}, \
  and\ \bibinfo {author} {\bibfnamefont {F.}~\bibnamefont {Nori}},\ }\href
  {\doibase 10.1103/PhysRevLett.120.140404} {\bibfield  {journal} {\bibinfo
  {journal} {Phys. Rev. Lett.}\ }\textbf {\bibinfo {volume} {120}},\ \bibinfo
  {pages} {140404} (\bibinfo {year} {2018})}\BibitemShut {NoStop}%
\bibitem [{\citenamefont {Guo}\ \emph {et~al.}(2017)\citenamefont {Guo},
  \citenamefont {Grimsmo}, \citenamefont {Kockum}, \citenamefont {Pletyukhov},\
  and\ \citenamefont {Johansson}}]{Guo-PRA2017}%
  \BibitemOpen
  \bibfield  {author} {\bibinfo {author} {\bibfnamefont {L.}~\bibnamefont
  {Guo}}, \bibinfo {author} {\bibfnamefont {A.}~\bibnamefont {Grimsmo}},
  \bibinfo {author} {\bibfnamefont {A.~F.}\ \bibnamefont {Kockum}}, \bibinfo
  {author} {\bibfnamefont {M.}~\bibnamefont {Pletyukhov}}, \ and\ \bibinfo
  {author} {\bibfnamefont {G.}~\bibnamefont {Johansson}},\ }\href {\doibase
  10.1103/PhysRevA.95.053821} {\bibfield  {journal} {\bibinfo  {journal} {Phys.
  Rev. A}\ }\textbf {\bibinfo {volume} {95}},\ \bibinfo {pages} {053821}
  (\bibinfo {year} {2017})}\BibitemShut {NoStop}%
\bibitem [{\citenamefont {Guo}\ \emph {et~al.}(2020{\natexlab{a}})\citenamefont
  {Guo}, \citenamefont {Kockum}, \citenamefont {Marquardt},\ and\ \citenamefont
  {Johansson}}]{Guo-PRR2020}%
  \BibitemOpen
  \bibfield  {author} {\bibinfo {author} {\bibfnamefont {L.}~\bibnamefont
  {Guo}}, \bibinfo {author} {\bibfnamefont {A.~F.}\ \bibnamefont {Kockum}},
  \bibinfo {author} {\bibfnamefont {F.}~\bibnamefont {Marquardt}}, \ and\
  \bibinfo {author} {\bibfnamefont {G.}~\bibnamefont {Johansson}},\ }\href
  {\doibase 10.1103/PhysRevResearch.2.043014} {\bibfield  {journal} {\bibinfo
  {journal} {Phys. Rev. Research}\ }\textbf {\bibinfo {volume} {2}},\ \bibinfo
  {pages} {043014} (\bibinfo {year} {2020}{\natexlab{a}})}\BibitemShut
  {NoStop}%
\bibitem [{\citenamefont {Guo}\ \emph {et~al.}(2020{\natexlab{b}})\citenamefont
  {Guo}, \citenamefont {Wang}, \citenamefont {Purdy},\ and\ \citenamefont
  {Taylor}}]{Guo-PRA2020}%
  \BibitemOpen
  \bibfield  {author} {\bibinfo {author} {\bibfnamefont {S.}~\bibnamefont
  {Guo}}, \bibinfo {author} {\bibfnamefont {Y.}~\bibnamefont {Wang}}, \bibinfo
  {author} {\bibfnamefont {T.}~\bibnamefont {Purdy}}, \ and\ \bibinfo {author}
  {\bibfnamefont {J.}~\bibnamefont {Taylor}},\ }\href {\doibase
  10.1103/PhysRevA.102.033706} {\bibfield  {journal} {\bibinfo  {journal}
  {Phys. Rev. A}\ }\textbf {\bibinfo {volume} {102}},\ \bibinfo {pages}
  {033706} (\bibinfo {year} {2020}{\natexlab{b}})}\BibitemShut {NoStop}%
\bibitem [{\citenamefont {Gonz\'alez-Tudela}\ \emph {et~al.}(2019)\citenamefont
  {Gonz\'alez-Tudela}, \citenamefont {Mu\~noz},\ and\ \citenamefont
  {Cirac}}]{Tudela-PRL2019}%
  \BibitemOpen
  \bibfield  {author} {\bibinfo {author} {\bibfnamefont {A.}~\bibnamefont
  {Gonz\'alez-Tudela}}, \bibinfo {author} {\bibfnamefont {C.~S.}\ \bibnamefont
  {Mu\~noz}}, \ and\ \bibinfo {author} {\bibfnamefont {J.~I.}\ \bibnamefont
  {Cirac}},\ }\href {\doibase 10.1103/PhysRevLett.122.203603} {\bibfield
  {journal} {\bibinfo  {journal} {Phys. Rev. Lett.}\ }\textbf {\bibinfo
  {volume} {122}},\ \bibinfo {pages} {203603} (\bibinfo {year}
  {2019})}\BibitemShut {NoStop}%
\bibitem [{\citenamefont {Zhao}\ and\ \citenamefont
  {Wang}(2020)}]{Zhao-PRA2020}%
  \BibitemOpen
  \bibfield  {author} {\bibinfo {author} {\bibfnamefont {W.}~\bibnamefont
  {Zhao}}\ and\ \bibinfo {author} {\bibfnamefont {Z.}~\bibnamefont {Wang}},\
  }\href {\doibase 10.1103/PhysRevA.101.053855} {\bibfield  {journal} {\bibinfo
   {journal} {Phys. Rev. A}\ }\textbf {\bibinfo {volume} {101}},\ \bibinfo
  {pages} {053855} (\bibinfo {year} {2020})}\BibitemShut {NoStop}%
\bibitem [{\citenamefont {Cheng}\ \emph {et~al.}()\citenamefont {Cheng},
  \citenamefont {Wang},\ and\ \citenamefont {Liu}}]{Cheng-Arxiv2021}%
  \BibitemOpen
  \bibfield  {author} {\bibinfo {author} {\bibfnamefont {W.}~\bibnamefont
  {Cheng}}, \bibinfo {author} {\bibfnamefont {Z.}~\bibnamefont {Wang}}, \ and\
  \bibinfo {author} {\bibfnamefont {Y.-x.}\ \bibnamefont {Liu}},\ }\href@noop
  {} {}\Eprint {http://arxiv.org/abs/2103.04542} {arXiv:2103.04542}
  \BibitemShut {NoStop}%
\bibitem [{\citenamefont {Astafiev}\ \emph {et~al.}(2010)\citenamefont
  {Astafiev}, \citenamefont {Zagoskin}, \citenamefont {Abdumalikov},
  \citenamefont {Pashkin}, \citenamefont {Yamamoto}, \citenamefont {Inomata},
  \citenamefont {Nakamura},\ and\ \citenamefont {Tsai}}]{Astafiev-Science2010}%
  \BibitemOpen
  \bibfield  {author} {\bibinfo {author} {\bibfnamefont {O.}~\bibnamefont
  {Astafiev}}, \bibinfo {author} {\bibfnamefont {A.~M.}\ \bibnamefont
  {Zagoskin}}, \bibinfo {author} {\bibfnamefont {A.~A.}\ \bibnamefont
  {Abdumalikov}}, \bibinfo {author} {\bibfnamefont {Y.~A.}\ \bibnamefont
  {Pashkin}}, \bibinfo {author} {\bibfnamefont {T.}~\bibnamefont {Yamamoto}},
  \bibinfo {author} {\bibfnamefont {K.}~\bibnamefont {Inomata}}, \bibinfo
  {author} {\bibfnamefont {Y.}~\bibnamefont {Nakamura}}, \ and\ \bibinfo
  {author} {\bibfnamefont {J.~S.}\ \bibnamefont {Tsai}},\ }\href {\doibase
  10.1126/science.1181918} {\bibfield  {journal} {\bibinfo  {journal}
  {Science}\ }\textbf {\bibinfo {volume} {327}},\ \bibinfo {pages} {840}
  (\bibinfo {year} {2010})}\BibitemShut {NoStop}%
\bibitem [{\citenamefont {Zhu}\ and\ \citenamefont {Jia}(2019)}]{Zhu-PRA2019}%
  \BibitemOpen
  \bibfield  {author} {\bibinfo {author} {\bibfnamefont {Y.~T.}\ \bibnamefont
  {Zhu}}\ and\ \bibinfo {author} {\bibfnamefont {W.~Z.}\ \bibnamefont {Jia}},\
  }\href {\doibase 10.1103/PhysRevA.99.063815} {\bibfield  {journal} {\bibinfo
  {journal} {Phys. Rev. A}\ }\textbf {\bibinfo {volume} {99}},\ \bibinfo
  {pages} {063815} (\bibinfo {year} {2019})}\BibitemShut {NoStop}%
\bibitem [{\citenamefont {Ask}\ \emph {et~al.}()\citenamefont {Ask},
  \citenamefont {Fang},\ and\ \citenamefont {Kockum}}]{Ask-Arxiv2020}%
  \BibitemOpen
  \bibfield  {author} {\bibinfo {author} {\bibfnamefont {A.}~\bibnamefont
  {Ask}}, \bibinfo {author} {\bibfnamefont {Y.-L.~L.}\ \bibnamefont {Fang}}, \
  and\ \bibinfo {author} {\bibfnamefont {A.~F.}\ \bibnamefont {Kockum}},\
  }\href@noop {} {}\Eprint {http://arxiv.org/abs/2011.15077} {arXiv:2011.15077}
  \BibitemShut {NoStop}%
\bibitem [{\citenamefont {Shen}\ and\ \citenamefont
  {Fan}(2005)}]{Shen-PRL2005}%
  \BibitemOpen
  \bibfield  {author} {\bibinfo {author} {\bibfnamefont {J.-T.}\ \bibnamefont
  {Shen}}\ and\ \bibinfo {author} {\bibfnamefont {S.}~\bibnamefont {Fan}},\
  }\href {\doibase 10.1103/PhysRevLett.95.213001} {\bibfield  {journal}
  {\bibinfo  {journal} {Phys. Rev. Lett.}\ }\textbf {\bibinfo {volume} {95}},\
  \bibinfo {pages} {213001} (\bibinfo {year} {2005})}\BibitemShut {NoStop}%
\bibitem [{\citenamefont {Ask}\ \emph {et~al.}(2019)\citenamefont {Ask},
  \citenamefont {Ekstr\"om}, \citenamefont {Delsing},\ and\ \citenamefont
  {Johansson}}]{Ask-PRA2019}%
  \BibitemOpen
  \bibfield  {author} {\bibinfo {author} {\bibfnamefont {A.}~\bibnamefont
  {Ask}}, \bibinfo {author} {\bibfnamefont {M.}~\bibnamefont {Ekstr\"om}},
  \bibinfo {author} {\bibfnamefont {P.}~\bibnamefont {Delsing}}, \ and\
  \bibinfo {author} {\bibfnamefont {G.}~\bibnamefont {Johansson}},\ }\href
  {\doibase 10.1103/PhysRevA.99.013840} {\bibfield  {journal} {\bibinfo
  {journal} {Phys. Rev. A}\ }\textbf {\bibinfo {volume} {99}},\ \bibinfo
  {pages} {013840} (\bibinfo {year} {2019})}\BibitemShut {NoStop}%
\bibitem [{\citenamefont {Dorner}\ and\ \citenamefont
  {Zoller}(2002)}]{Dorner-PRA2002}%
  \BibitemOpen
  \bibfield  {author} {\bibinfo {author} {\bibfnamefont {U.}~\bibnamefont
  {Dorner}}\ and\ \bibinfo {author} {\bibfnamefont {P.}~\bibnamefont
  {Zoller}},\ }\href {\doibase 10.1103/PhysRevA.66.023816} {\bibfield
  {journal} {\bibinfo  {journal} {Phys. Rev. A}\ }\textbf {\bibinfo {volume}
  {66}},\ \bibinfo {pages} {023816} (\bibinfo {year} {2002})}\BibitemShut
  {NoStop}%
\bibitem [{\citenamefont {Tufarelli}\ \emph {et~al.}(2013)\citenamefont
  {Tufarelli}, \citenamefont {Ciccarello},\ and\ \citenamefont
  {Kim}}]{Tufarelli-PRA2013}%
  \BibitemOpen
  \bibfield  {author} {\bibinfo {author} {\bibfnamefont {T.}~\bibnamefont
  {Tufarelli}}, \bibinfo {author} {\bibfnamefont {F.}~\bibnamefont
  {Ciccarello}}, \ and\ \bibinfo {author} {\bibfnamefont {M.~S.}\ \bibnamefont
  {Kim}},\ }\href {\doibase 10.1103/PhysRevA.87.013820} {\bibfield  {journal}
  {\bibinfo  {journal} {Phys. Rev. A}\ }\textbf {\bibinfo {volume} {87}},\
  \bibinfo {pages} {013820} (\bibinfo {year} {2013})}\BibitemShut {NoStop}%
\bibitem [{\citenamefont {Pichler}\ and\ \citenamefont
  {Zoller}(2016)}]{Pichler-PRL2016}%
  \BibitemOpen
  \bibfield  {author} {\bibinfo {author} {\bibfnamefont {H.}~\bibnamefont
  {Pichler}}\ and\ \bibinfo {author} {\bibfnamefont {P.}~\bibnamefont
  {Zoller}},\ }\href {\doibase 10.1103/PhysRevLett.116.093601} {\bibfield
  {journal} {\bibinfo  {journal} {Phys. Rev. Lett.}\ }\textbf {\bibinfo
  {volume} {116}},\ \bibinfo {pages} {093601} (\bibinfo {year}
  {2016})}\BibitemShut {NoStop}%
\bibitem [{\citenamefont {Guimond}\ \emph {et~al.}(2016)\citenamefont
  {Guimond}, \citenamefont {Roulet}, \citenamefont {Le},\ and\ \citenamefont
  {Scarani}}]{Guimond-PRA2016}%
  \BibitemOpen
  \bibfield  {author} {\bibinfo {author} {\bibfnamefont {P.-O.}\ \bibnamefont
  {Guimond}}, \bibinfo {author} {\bibfnamefont {A.}~\bibnamefont {Roulet}},
  \bibinfo {author} {\bibfnamefont {H.~N.}\ \bibnamefont {Le}}, \ and\ \bibinfo
  {author} {\bibfnamefont {V.}~\bibnamefont {Scarani}},\ }\href {\doibase
  10.1103/PhysRevA.93.023808} {\bibfield  {journal} {\bibinfo  {journal} {Phys.
  Rev. A}\ }\textbf {\bibinfo {volume} {93}},\ \bibinfo {pages} {023808}
  (\bibinfo {year} {2016})}\BibitemShut {NoStop}%
\bibitem [{\citenamefont {Laakso}\ and\ \citenamefont
  {Pletyukhov}(2014)}]{Laakso-PRL2014}%
  \BibitemOpen
  \bibfield  {author} {\bibinfo {author} {\bibfnamefont {M.}~\bibnamefont
  {Laakso}}\ and\ \bibinfo {author} {\bibfnamefont {M.}~\bibnamefont
  {Pletyukhov}},\ }\href {\doibase 10.1103/PhysRevLett.113.183601} {\bibfield
  {journal} {\bibinfo  {journal} {Phys. Rev. Lett.}\ }\textbf {\bibinfo
  {volume} {113}},\ \bibinfo {pages} {183601} (\bibinfo {year}
  {2014})}\BibitemShut {NoStop}%
\bibitem [{\citenamefont {Fang}\ and\ \citenamefont
  {Baranger}(2015)}]{Fang-PRA2015}%
  \BibitemOpen
  \bibfield  {author} {\bibinfo {author} {\bibfnamefont {Y.-L.~L.}\
  \bibnamefont {Fang}}\ and\ \bibinfo {author} {\bibfnamefont {H.~U.}\
  \bibnamefont {Baranger}},\ }\href {\doibase 10.1103/PhysRevA.91.053845}
  {\bibfield  {journal} {\bibinfo  {journal} {Phys. Rev. A}\ }\textbf {\bibinfo
  {volume} {91}},\ \bibinfo {pages} {053845} (\bibinfo {year}
  {2015})}\BibitemShut {NoStop}%
\bibitem [{\citenamefont {Milonni}\ and\ \citenamefont
  {Knight}(1974)}]{Milonni-PRA1974}%
  \BibitemOpen
  \bibfield  {author} {\bibinfo {author} {\bibfnamefont {P.~W.}\ \bibnamefont
  {Milonni}}\ and\ \bibinfo {author} {\bibfnamefont {P.~L.}\ \bibnamefont
  {Knight}},\ }\href {\doibase 10.1103/PhysRevA.10.1096} {\bibfield  {journal}
  {\bibinfo  {journal} {Phys. Rev. A}\ }\textbf {\bibinfo {volume} {10}},\
  \bibinfo {pages} {1096} (\bibinfo {year} {1974})}\BibitemShut {NoStop}%
\bibitem [{\citenamefont {Rist}\ \emph {et~al.}(2008)\citenamefont {Rist},
  \citenamefont {Eschner}, \citenamefont {Hennrich},\ and\ \citenamefont
  {Morigi}}]{Rist-PRA2008}%
  \BibitemOpen
  \bibfield  {author} {\bibinfo {author} {\bibfnamefont {S.}~\bibnamefont
  {Rist}}, \bibinfo {author} {\bibfnamefont {J.}~\bibnamefont {Eschner}},
  \bibinfo {author} {\bibfnamefont {M.}~\bibnamefont {Hennrich}}, \ and\
  \bibinfo {author} {\bibfnamefont {G.}~\bibnamefont {Morigi}},\ }\href
  {\doibase 10.1103/PhysRevA.78.013808} {\bibfield  {journal} {\bibinfo
  {journal} {Phys. Rev. A}\ }\textbf {\bibinfo {volume} {78}},\ \bibinfo
  {pages} {013808} (\bibinfo {year} {2008})}\BibitemShut {NoStop}%
\bibitem [{\citenamefont {van Loo}\ \emph {et~al.}(2013)\citenamefont {van
  Loo}, \citenamefont {Fedorov}, \citenamefont {Lalumi{\`e}re}, \citenamefont
  {Sanders}, \citenamefont {Blais},\ and\ \citenamefont
  {Wallraff}}]{Loo-Science2013}%
  \BibitemOpen
  \bibfield  {author} {\bibinfo {author} {\bibfnamefont {A.~F.}\ \bibnamefont
  {van Loo}}, \bibinfo {author} {\bibfnamefont {A.}~\bibnamefont {Fedorov}},
  \bibinfo {author} {\bibfnamefont {K.}~\bibnamefont {Lalumi{\`e}re}}, \bibinfo
  {author} {\bibfnamefont {B.~C.}\ \bibnamefont {Sanders}}, \bibinfo {author}
  {\bibfnamefont {A.}~\bibnamefont {Blais}}, \ and\ \bibinfo {author}
  {\bibfnamefont {A.}~\bibnamefont {Wallraff}},\ }\href {\doibase
  10.1126/science.1244324} {\bibfield  {journal} {\bibinfo  {journal}
  {Science}\ }\textbf {\bibinfo {volume} {342}},\ \bibinfo {pages} {1494}
  (\bibinfo {year} {2013})}\BibitemShut {NoStop}%
\bibitem [{\citenamefont {Fang}\ \emph {et~al.}(2014)\citenamefont {Fang},
  \citenamefont {Zheng},\ and\ \citenamefont {Baranger}}]{Fang-RPJ2014}%
  \BibitemOpen
  \bibfield  {author} {\bibinfo {author} {\bibfnamefont {Y.-L.~L.}\
  \bibnamefont {Fang}}, \bibinfo {author} {\bibfnamefont {H.}~\bibnamefont
  {Zheng}}, \ and\ \bibinfo {author} {\bibfnamefont {H.~U.}\ \bibnamefont
  {Baranger}},\ }\href {https://doi.org/10.1140/epjqt3} {\bibfield  {journal}
  {\bibinfo  {journal} {EPJ Quantum Technology}\ }\textbf {\bibinfo {volume}
  {1}} (\bibinfo {year} {2014})}\BibitemShut {NoStop}%
\bibitem [{\citenamefont {Hoi}\ \emph {et~al.}(2011)\citenamefont {Hoi},
  \citenamefont {Wilson}, \citenamefont {Johansson}, \citenamefont {Palomaki},
  \citenamefont {Peropadre},\ and\ \citenamefont {Delsing}}]{Hoi-PRL2011}%
  \BibitemOpen
  \bibfield  {author} {\bibinfo {author} {\bibfnamefont {I.-C.}\ \bibnamefont
  {Hoi}}, \bibinfo {author} {\bibfnamefont {C.~M.}\ \bibnamefont {Wilson}},
  \bibinfo {author} {\bibfnamefont {G.}~\bibnamefont {Johansson}}, \bibinfo
  {author} {\bibfnamefont {T.}~\bibnamefont {Palomaki}}, \bibinfo {author}
  {\bibfnamefont {B.}~\bibnamefont {Peropadre}}, \ and\ \bibinfo {author}
  {\bibfnamefont {P.}~\bibnamefont {Delsing}},\ }\href {\doibase
  10.1103/PhysRevLett.107.073601} {\bibfield  {journal} {\bibinfo  {journal}
  {Phys. Rev. Lett.}\ }\textbf {\bibinfo {volume} {107}},\ \bibinfo {pages}
  {073601} (\bibinfo {year} {2011})}\BibitemShut {NoStop}%
\bibitem [{\citenamefont {Chang}\ \emph {et~al.}(2012)\citenamefont {Chang},
  \citenamefont {Jiang}, \citenamefont {Gorshkov},\ and\ \citenamefont
  {Kimble}}]{Chang-NJP2012}%
  \BibitemOpen
  \bibfield  {author} {\bibinfo {author} {\bibfnamefont {D.~E.}\ \bibnamefont
  {Chang}}, \bibinfo {author} {\bibfnamefont {L.}~\bibnamefont {Jiang}},
  \bibinfo {author} {\bibfnamefont {A.~V.}\ \bibnamefont {Gorshkov}}, \ and\
  \bibinfo {author} {\bibfnamefont {H.~J.}\ \bibnamefont {Kimble}},\ }\href
  {\doibase 10.1088/1367-2630/14/6/063003} {\bibfield  {journal} {\bibinfo
  {journal} {New Journal of Physics}\ }\textbf {\bibinfo {volume} {14}},\
  \bibinfo {pages} {063003} (\bibinfo {year} {2012})}\BibitemShut {NoStop}%
\bibitem [{\citenamefont {Greenberg}\ \emph {et~al.}(2021)\citenamefont
  {Greenberg}, \citenamefont {Shtygashev},\ and\ \citenamefont
  {Moiseev}}]{Greenberg-PRA2021}%
  \BibitemOpen
  \bibfield  {author} {\bibinfo {author} {\bibfnamefont {Y.~S.}\ \bibnamefont
  {Greenberg}}, \bibinfo {author} {\bibfnamefont {A.~A.}\ \bibnamefont
  {Shtygashev}}, \ and\ \bibinfo {author} {\bibfnamefont {A.~G.}\ \bibnamefont
  {Moiseev}},\ }\href {\doibase 10.1103/PhysRevA.103.023508} {\bibfield
  {journal} {\bibinfo  {journal} {Phys. Rev. A}\ }\textbf {\bibinfo {volume}
  {103}},\ \bibinfo {pages} {023508} (\bibinfo {year} {2021})}\BibitemShut
  {NoStop}%
\bibitem [{\citenamefont {Mukhopadhyay}\ and\ \citenamefont
  {Agarwal}(2020)}]{Mukhopadhyay-PRA2020}%
  \BibitemOpen
  \bibfield  {author} {\bibinfo {author} {\bibfnamefont {D.}~\bibnamefont
  {Mukhopadhyay}}\ and\ \bibinfo {author} {\bibfnamefont {G.~S.}\ \bibnamefont
  {Agarwal}},\ }\href {\doibase 10.1103/PhysRevA.101.063814} {\bibfield
  {journal} {\bibinfo  {journal} {Phys. Rev. A}\ }\textbf {\bibinfo {volume}
  {101}},\ \bibinfo {pages} {063814} (\bibinfo {year} {2020})}\BibitemShut
  {NoStop}%
\bibitem [{\citenamefont {Poudyal}\ and\ \citenamefont
  {Mirza}(2020)}]{Poudyal-PRR2020}%
  \BibitemOpen
  \bibfield  {author} {\bibinfo {author} {\bibfnamefont {B.}~\bibnamefont
  {Poudyal}}\ and\ \bibinfo {author} {\bibfnamefont {I.~M.}\ \bibnamefont
  {Mirza}},\ }\href {\doibase 10.1103/PhysRevResearch.2.043048} {\bibfield
  {journal} {\bibinfo  {journal} {Phys. Rev. Research}\ }\textbf {\bibinfo
  {volume} {2}},\ \bibinfo {pages} {043048} (\bibinfo {year}
  {2020})}\BibitemShut {NoStop}%
\end{thebibliography}%

\end{document}